\documentclass[twocolumn, pre,floatfix]{revtex4}
\usepackage{eurosym}
\usepackage{graphicx,amssymb,amstext,amsmath}
\usepackage{epsfig}
\usepackage[colorlinks = true,
            linkcolor = blue,
            urlcolor  = blue,
            citecolor = blue,
            anchorcolor = blue]{hyperref}
\usepackage{epstopdf}
\usepackage[english]{babel}
\usepackage[autostyle]{csquotes}
\usepackage{lipsum}
\usepackage{subfigure,url}
\usepackage{color,graphics}
\usepackage{braket}
\usepackage{float}
\usepackage{mathtools}
\usepackage[pass,paperwidth=13in in,paperheight=11in]{geometry}
\draft
\newcommand*{\addheight}[2][.5ex]{%
  \raisebox{0pt}[\dimexpr\height+(#1)\relax]{#2}%
}
\begin{document}

\title{Randomizing Quantum Walk} %Title of paper

\author{Rashid Ahmad, Safia Bibi, Uzma Sajjad}

\affiliation{Department of Physics, Kohat University of Science and Technology, Kohat 26000, Khyber-Pakhtunkhwa, Pakistan}

\date{\today}

%\pacs{52.25.Xz, 52.27.Ep, 52.35.Mw, 52.35.Sb}
%\pacs{52.25.Xz, 52.30.Ex, 52.35.-g}% insert suggested PACS numbers in braces on next line
\begin{abstract}
The evolution of a walker in standard "Discrete-time Quantum Walk (DTQW)" is determined by coin and shift unitary operators. The conditional shift operator shifts the position of the walker to right or left by unit step size while the direction of motion is specified by the coin operator. This scenario can be generalized by choosing the step size randomly at each step in some specific interval. For example, the value of the roll of a dice can be used to specify the step size after throwing the coin. Let us call such a quantum walk  "Discrete-time Random Step Quantum Walk (DTRSQW)". A completely random probability distribution is obtained whenever the walker follows the DTRSQW.  We have also analyzed two more types of quantum walks, the "Discrete-time Un-biased Quantum Walk (DTUBQW)"  and the "Discrete-time Biased Quantum Walk (DTBQW)". In the first type, the step size is kept different than unit size but the same for left and right shifts, whereas in the second type left and right shifts can also be different. The probability distribution in DTUBQW is found to follow a certain rule.  The standard deviation ($\sigma$) of DTRSQW is higher than DTQW and hence DTRSQW spreads faster. The $\sigma$ of DTUBQW  shows sawtooth behavior with faster spread than DTQW for some specific values of rotation angles and steps.
\end{abstract}
\maketitle %\maketitle must follow title, authors, abstract and \pacs


\section{Introduction}
Walks are a rudimentary procedure that is composed of a series of steps that shift the walker position. If the steps are randomly taken the final position of the walker is randomized and such a walk is called a random walk. They can be studied in terms of classical mechanics because they follow the Liouville equation. To model, a physical system one can use random walks as a fundamental mathematical tool. For example, they can be used to solve genetic draft in population genetics, the Brownian motion of molecules  \cite{brownian} and to find out the solution of Laplace equation \cite{diff}. Random walks have also proved to be quite useful in computational sciences e.g. to calculate probabilistic turning machines \cite{slutsky}. But there are also some walks in nature, which are not according to the definition of classical mechanics, e.g. the propagation of a single excitation in a crystal, during photosynthesis the effective energy transport in plants or the transport of quantum information on quantum networks. To model, such phenomena walks are to be extended to the quantum domain known as quantum walks. In fact, most quantum processes can be viewed as generalized quantum walks.  Since their introduction, quantum walks have gained considerable attention due to their ability to model physical systems of diverse nature. Several aspects of quantum walks are discovered and the straightforward construction of quantum walks makes them suitable tools for simulating other quantum systems \cite{kempe, andracas}.

\par Like classical walks, quantum walks are also being researched on a large scale. However, most commonly studied types are "Continuous-time Quantum Walk (CTQW)" and "Discrete-time Quantum Walk (DTQW)". The CTQW was introduced in 1998 by Farhi and Gutmann. In such a walk one can directly determine the speed in the position space \cite{cont1,cont2,cont3}. While the DTQW was introduced by Watrous in 2001 \cite{disc1,disc2,disc3} its specific form controlled by Hadamard coin introduced by Ambainis et al \cite{Ambainis}. got more attention due to its simplicity.
\par The results of a quantum walk with continuous-time and with discrete-time are often comparable, but the coin of freedom has shown that the discrete time-variant in a given context is stronger than others. The quantum walk with discrete-time is also a structure that is very similar to its classical counterpart. A coin flip is replaced by a quantum coin operation that determines the superposition of the direction in which the particle moves simultaneously. The quantum extraction activity followed by the unitary shift operation is repeated without using intermediate measurements to implement a large number of steps.
\par DTQWs have vast applications, e.g. in advance technology, they are used to develop fast algorithms \cite{search1,search2,speedup,algorithm1,algorithm2,algorithm3,algorithm4} for quantum computation tasks. DTQWs are also used for the simulation of the topological phenomenon in condensed matter systems \cite{topo1,topo2,topo3,chiral}. By these applications researchers are motivated to study the behavior of quantum walks \cite{dim,graphs,mcoin,mparticle,aperiodic,decoherent,history}. In the conditional DTQW step size is usually taken as a unit and fixed for each time the coin is tossed. Quantum walks with step size different than unit size are considered before. In \cite{biased} the stability of recurrence of quantum walk on a line is tested against bias. While the quantum walks with sequential aperiodic jumps are recently analyzed \cite{sequential}.

\par Here, we generalize this walk in three ways. In the first case, we introduce another degree of freedom by allowing the step size to be different than unit size and randomly chosen after each throw of the coin. We found that this procedure completely randomizes the DTQW. We call this type of walk as "Discrete-time Random Step Quantum Walk (DTRSQW)" as shown in Fig. \ref{fig:1} (a). The standard deviation $\sigma$ for this walk goes higher than standard DTQW. This feature can help in designing better algorithms and simulations as compared to DTQW. The second generalization is about choosing fixed step sizes which are unbiased shifts i.e. step size remains the same at each step. This quantum walk is named as "Discrete-time Un-biased Quantum Walk (DTUBQW)" as shown in Fig. \ref{fig:1} (b). For this type of walk probability distribution follow a certain rule i.e. the positions of the peaks can be determined with the help of a formula. The $\sigma$ shows a sawtooth behavior and goes higher than the DTQW for higher angles and small step size. In the third situation, the biased step is used for the different outcomes of the coin. We call this walk as "Discrete-time Biased Quantum Walk (DTBQW)" as shown in Fig. \ref{fig:1} (c). The standard deviation $\sigma$ gets larger values for Left Step Size (LSS) larger than Right Step Size (RSS).

Randomness in the quantum phenomenon is an important concept that finds many applications in fundamental quantum mechanics and quantum technology         \cite{Bera}. The DTRSQW can be useful in stimulating a variety of physical phenomena that are quantum and random simultaneously. The most important is the quantum Brownian motion. It will be interesting to apply DTRSQW to the quantum Brownian motion. Besides, such walks can also be realized in other systems for example in explaining the energy transfer in photosynthesis where excitation in the photosynthetic unit energy is supposed to be doing a random walk \cite{Sension,Hoyer}. On the other hand, it can open a new dimension in designing quantum algorithms especially search algorithms that are based on the DTRSQW.

\par This article is organized in the following way. In Section II, we introduce the DTRSQW in detail and the corresponding probability distribution. In Section III, the second variant of a quantum walk DTUBQW with an unbiased fixed step is discussed. The third type of walk,  DTBQW with unequal step size for different outcomes of the coin is introduced in Section IV. In Section V,  the standard deviation of all three walks is discussed. In section VI, we give conclusions of this work.
\onecolumngrid
\begin{center}
\noindent
\begin{figure}[H]
\begin{minipage}{0.98\columnwidth}
\begin{tabular}{ccc}
      \addheight {\includegraphics[width=57mm]{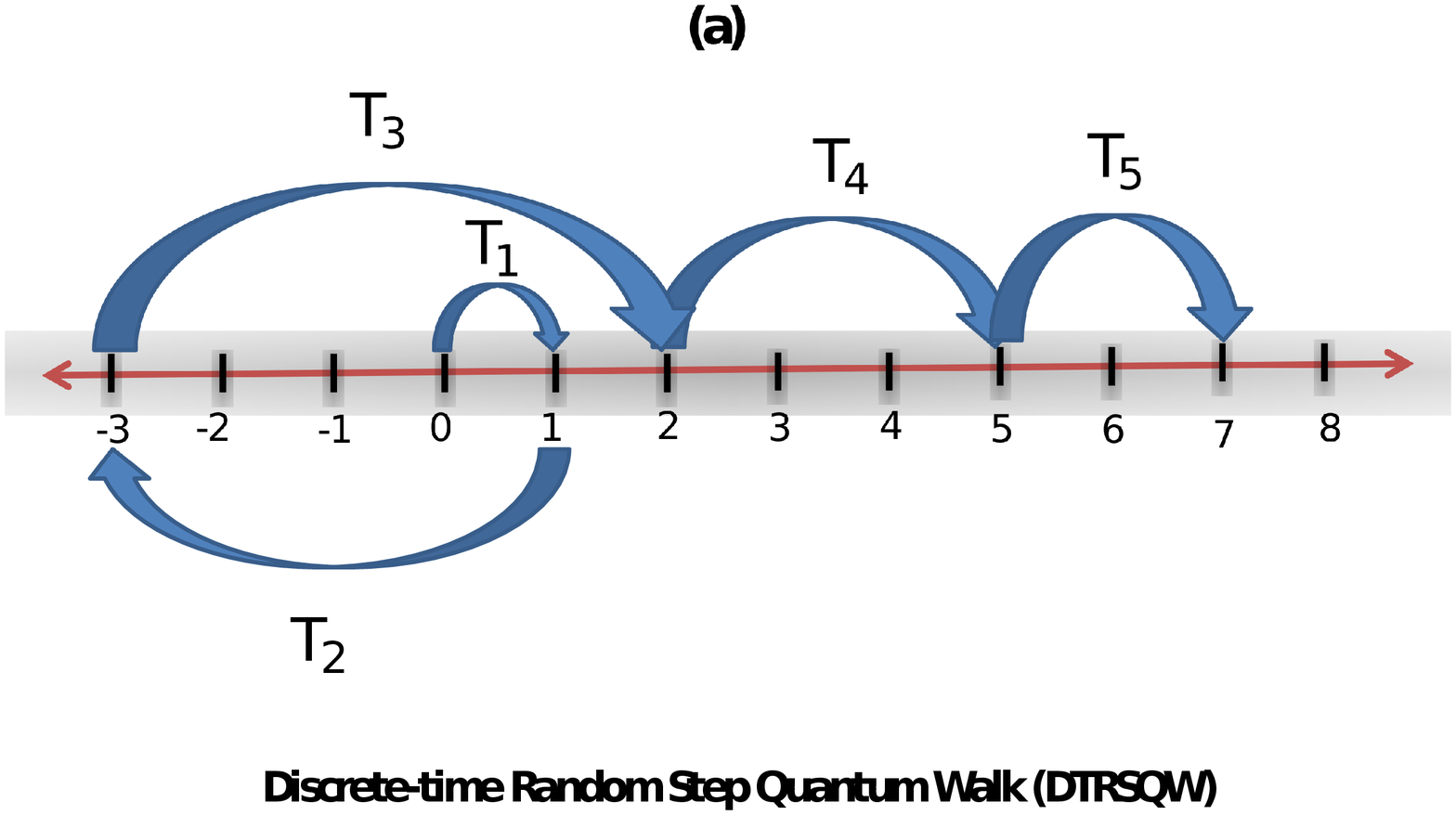}} &
      \addheight {\includegraphics[width=57mm]{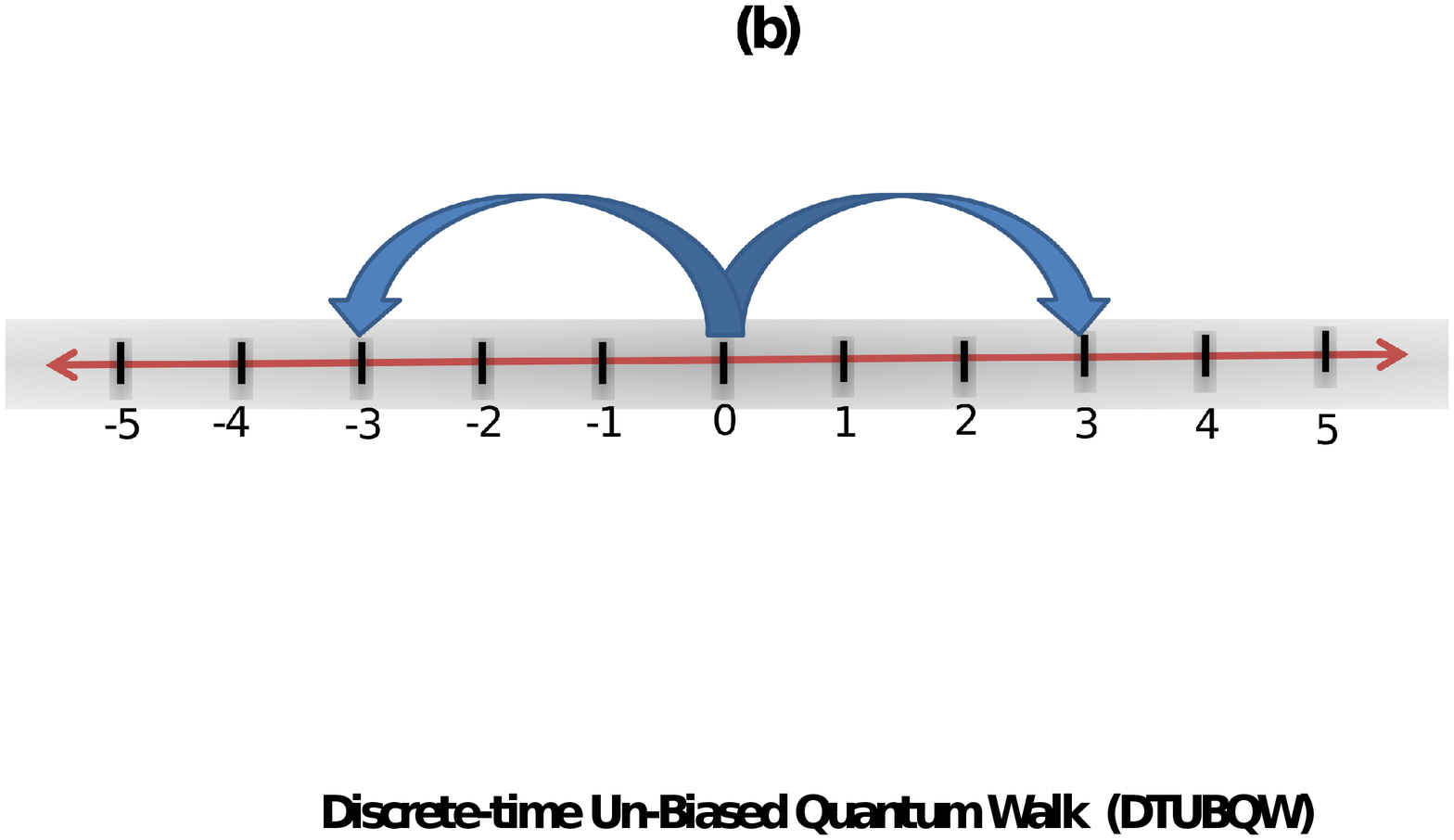}} &
      \addheight{\includegraphics[width=57mm]{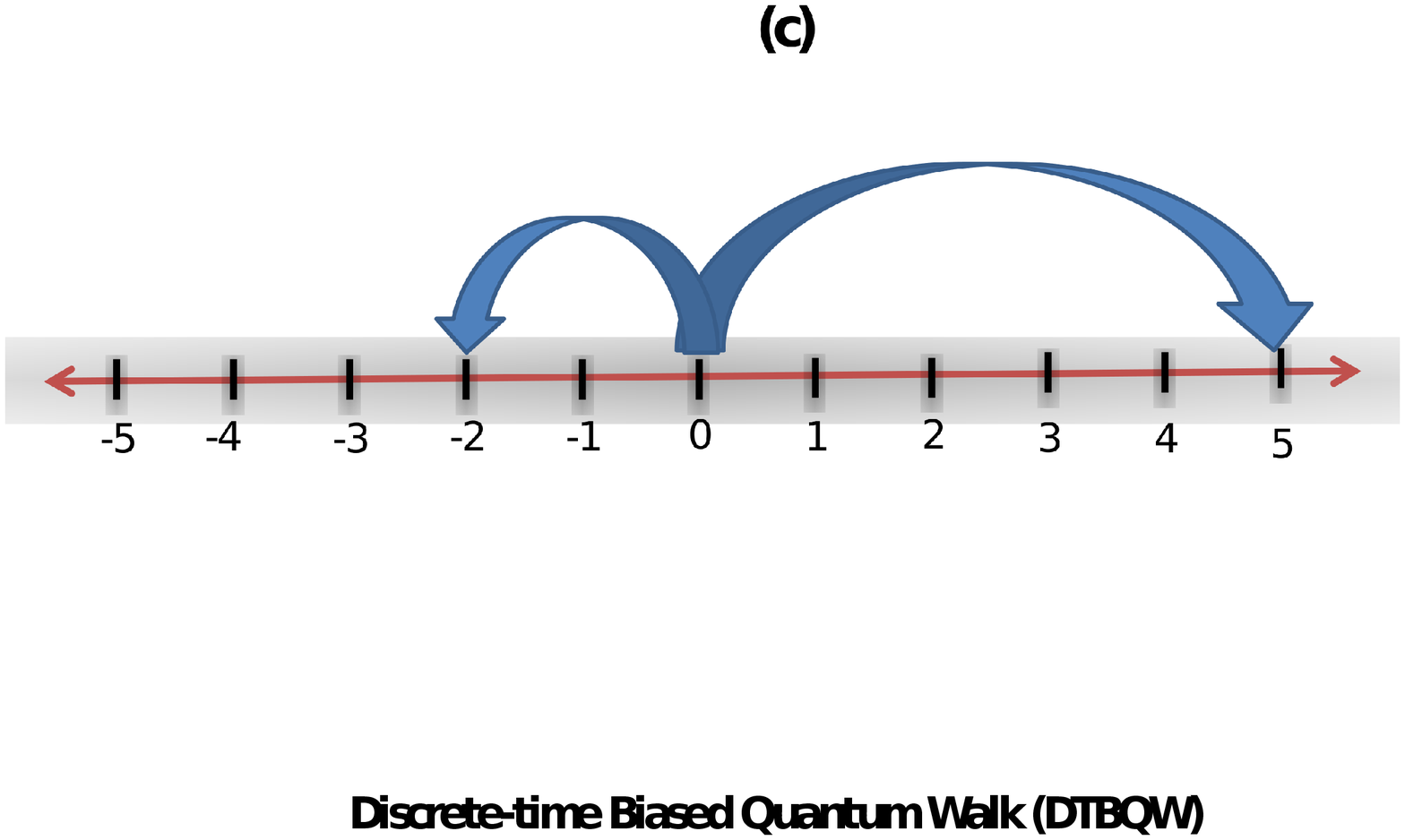}}\\
 \end{tabular}
\caption{Schematic  diagrams of walker that moves to either left or  right depending upon the outcome  of  a coin (a) DTRQW: the  step size is randomly chosen at each step.  (b)  DTRSQW:  equal step size that can be different than unit step size. (c)  DTRSQW: step size can be different for left and right movement.}
\label{fig:1}
\end{minipage}
\end{figure}
\end{center}
\twocolumngrid
\section{Discrete-time Random Quantum Walk (DTRSQW)}

In a DTQW, the application of two unitary operators a \enquote{coin operator} and a \enquote{conditional shift operator} determines the evolution of a walker. A coin operator acts on the internal state of a walker,  generally known as spin-up and spin-down states for a walker with two internal degrees of freedom. The shift operator is conditional and acts on the external degree of freedom and moves the walker either to the left or right depending on its internal state. Here, we introduce the interval for selecting the size of the step which randomizes the quantum walk. For a single-qubit walker, the coin (Hilbert) space, $\mathcal{H}_c$, is spanned by ${\ket 0, \ket 1}$. The position (Hilbert) space, $\mathcal{H}_p$, is spanned by ${\ket x: x \epsilon\:\mathbb{Z}}$. Thus the Hilbert space of the system is the tensor product of Hilbert spaces of the components, i.e. $\mathcal{H}_p \otimes \mathcal{H}_c$. The walker's internal degrees of freedom play an important role to have many features of DTQWs.

\par In DTRSQW we keep the coin operator of standard DTQW given in the following;

\begin{align}
\nonumber \hat{C}=&(\cos \theta \:  \ket 0_C \bra 0 +\sin \theta\: \ket 0_C \bra 1
\label{coin} \\+&\sin \theta \: \ket 1_C \bra 0 -\cos \theta\: \ket 1_C \bra 1_C),
\end{align}
here $\theta$ is the rotation angle.  The conditional shift operator however defined as follows

\begin{align}\notag
\hat{S}=&\ket 0_C \bra 0 \otimes \sum_x \ket {x-j}_P \bra x
\label{shift} \\+&\ket 1_C \bra 1 \otimes \sum_x \ket {x+j}_P \bra x,
\end{align}
This shift operator has an integer "j" that is randomly chosen from interval [1, n] at each step where n is an integer.   The walk operator, $\hat{U}=\hat{S}\hat{C}$, acts on the initial state ($\phi_{\text{int}}$) of the walker. The walk operator acts repeatedly to the initial states for many steps and leads to the evolution of the walk, i.e.

\begin{align} \label{initial1}
 \ket \phi_{\text{int1}}= &\ket 0_C \otimes \ket 0_P
 \\\ket \phi_{\text{int2}}= &\ket 1_C \otimes \ket 0_P
\end{align}
here $\phi_{\text{T}}$ is the final state after $T$ steps of the walk.
\par The states $\phi_{\text{int1}}(\phi_{\text{int2}})$ describes a particle prepared in spin-up (spin-down)  internal state which is spatially localized around the origin of the 1D lattice.

The stepwise evolution of the initial state $\phi_{\text{int1}}$ of the walker can be mathematically written as
\begin{align}
\notag \ket 0_C \otimes \ket 0_P &\xRightarrow[\text{}]{\text{1st}}\cos\theta\ket 0_C\otimes \ket{{{-x}_{[1, n]}}_P}
\\ \notag &+\sin\theta\ket 1_C\otimes \ket{{{x}_{[1, n]}}_P}
\\ \notag &\xRightarrow[\text{}]{\text{2nd}}\cos^{2}\theta \:\ket 0_C \otimes \ket{{{-x}_{[1, 2n]}}_P}
\\ \notag &+\sin^{2}\theta \:\ket 0_C\otimes \ket {{{-x}_{[1, 2n]}}_P}
\\ \notag &+\cos\theta \sin\theta \ket 1_C \otimes\ket {{{x}_{[1, 2n]}}_P}
\\ \notag &-\cos\theta \sin\theta \ket 1_C \otimes\ket {{{x}_{[1, 2n]}}_P}
\\ \label{f1} \xRightarrow[\text{}]{\text{Tth}} &\cos^{T}\theta    \: \ket 0_C \otimes\ket {{{x}_{[1, -Tn]}}_P}+.......,
\end{align}

For a standard DTQW, the shift operator moves the walker either to the left or right depending on the outcomes of the coin. The random number, however, can be chosen from a specific interval. In DTRQW the shift operator moves the walker randomly of the step size chosen form a specific interval.
In Fig. \ref{fig:a}  we have shown the probability distribution for angle $\theta=\pi/4$, in the top panel of Fig. \ref{fig:a} \big(a, b, c\big) the step size is chosen randomly from interval [1, 3]  and walker is shifted accordingly. At the next position, the size of the next step is again chosen randomly from the interval the same interval.  In the middle panel of Fig. \ref{fig:a} \big(d, e, f \big) step size is randomly selected from  the interval [1, 6] and similarly in the bottom panel of Fig. \ref{fig:a} \big(g, h, i\big)  the interval selected is [1, 10]. We have shown three runs for each interval to illustrate that in each run the probability distribution changes randomly and the quantum walk is completely randomized independent of the interval chosen.
\onecolumngrid
\begin{center}
\noindent
\begin{figure}[H]
\begin{minipage}{0.98\columnwidth}

\begin{tabular}{ccc}
      \addheight {\includegraphics[width=57mm]{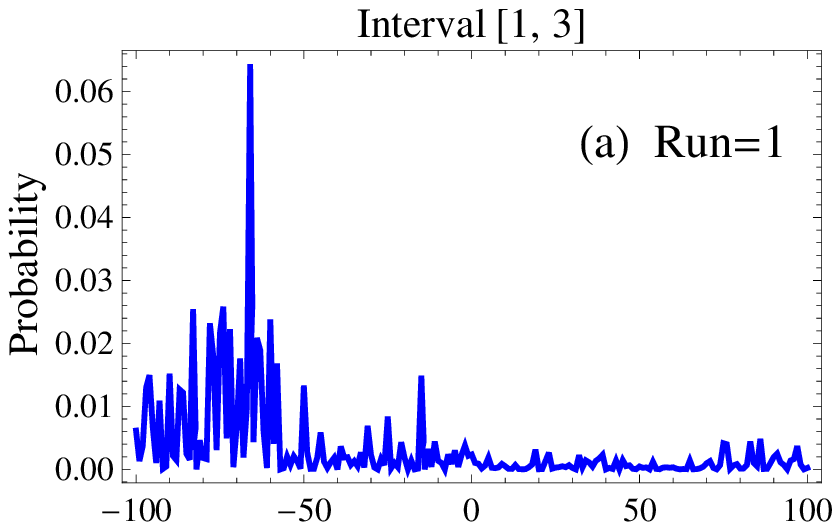}} &
      \addheight {\includegraphics[width=57mm]{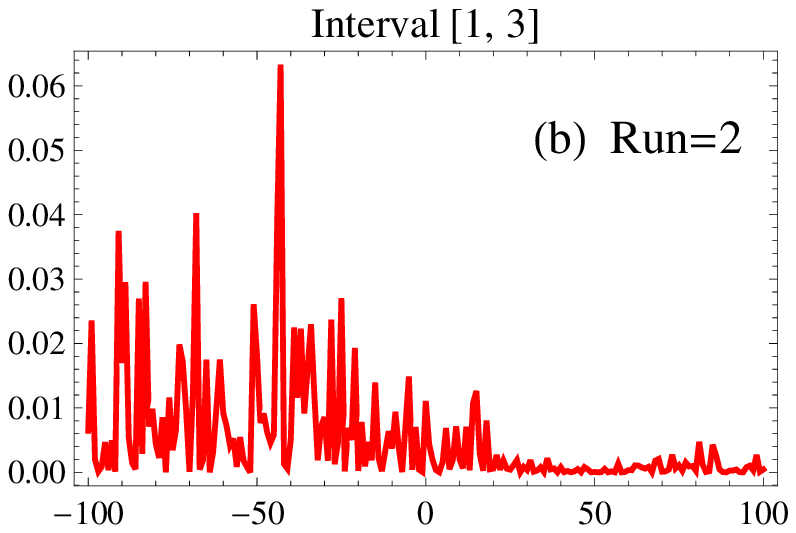}} &
      \addheight{\includegraphics[width=57mm]{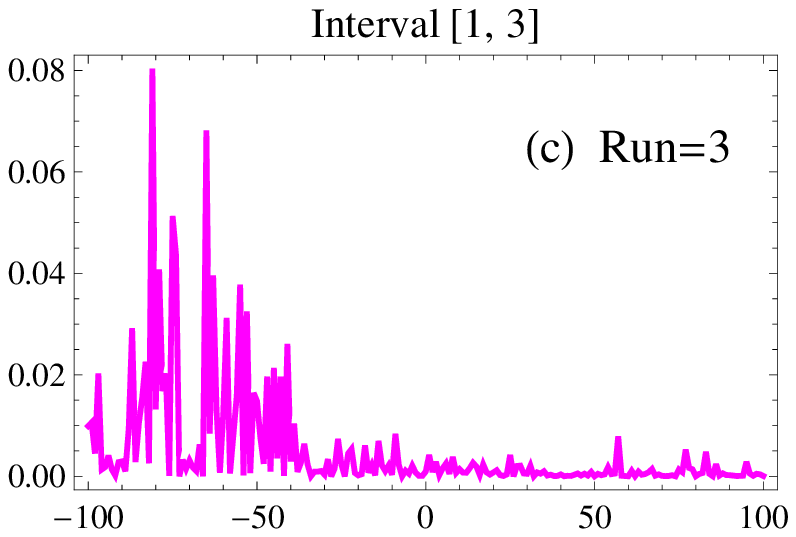}}\\
      \addheight{\includegraphics[width=57mm]{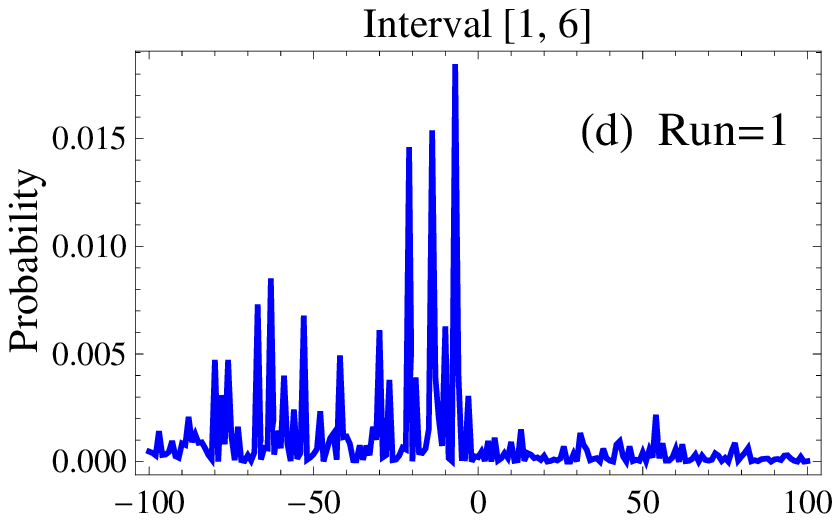}} &
      \addheight{\includegraphics[width=57mm]{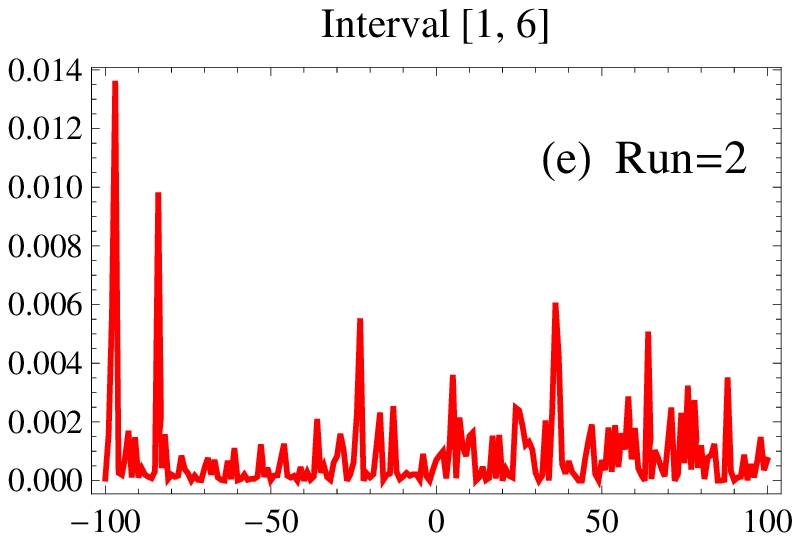}} &
      \addheight{\includegraphics[width=57mm]{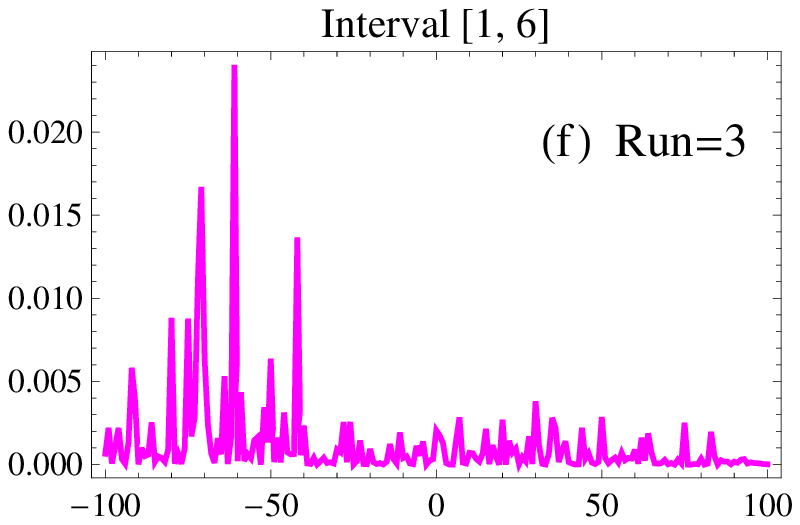}}\\
       \addheight{\includegraphics[width=57mm]{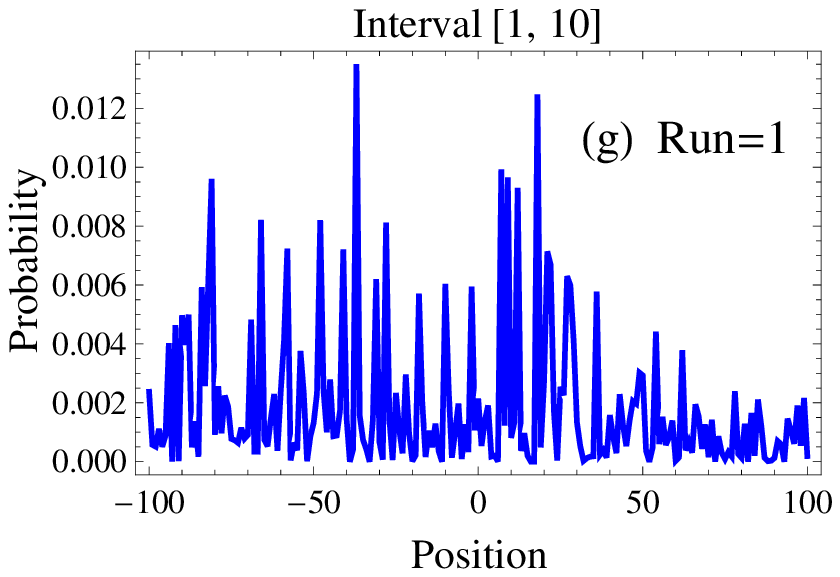}}&
        \addheight{\includegraphics[width=57mm]{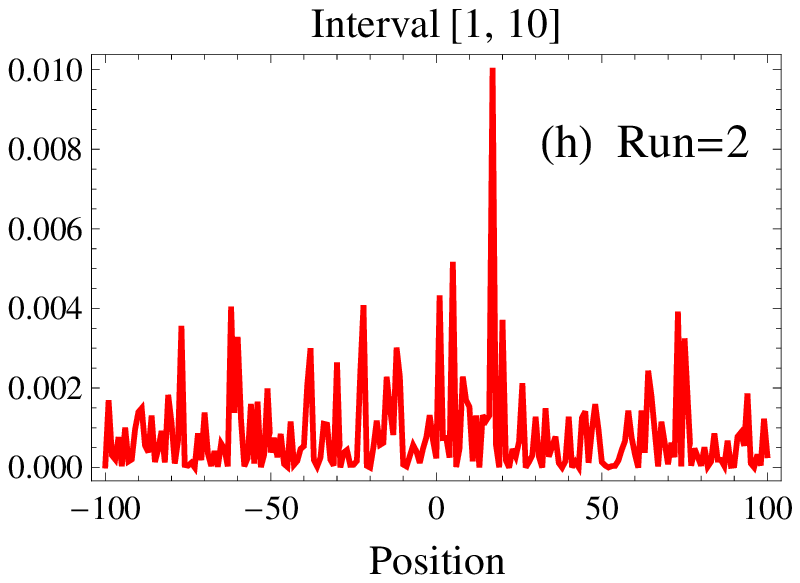}}&
        \addheight{\includegraphics[width=57mm]{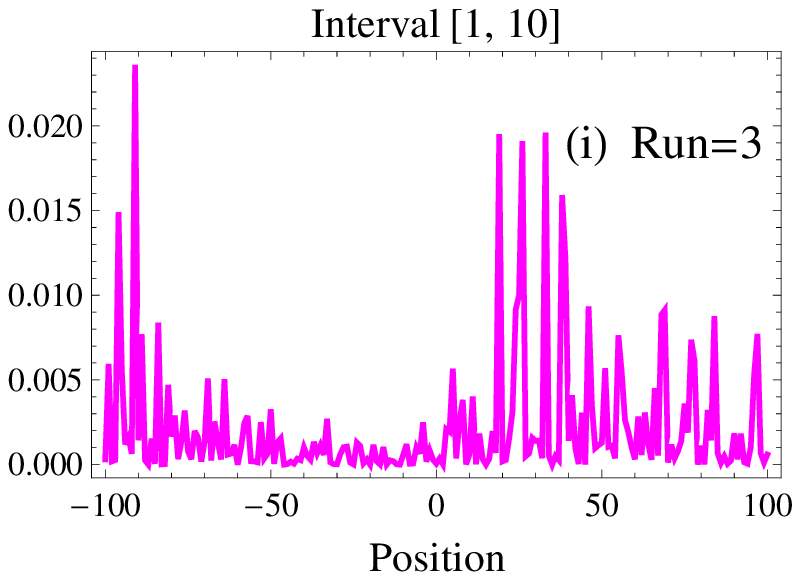}}\\
 \end{tabular}
\caption{Probability distributions of the DTRSQW after T $=100$ for rotation angle $(\theta=\pi/4)$. The top panel \big(Figs.(a, b, c)\big) shows the probability distributions for Run (1). The middle panel \big(Figs. (d, e, f)\big) shows the probability distributions for Run (2). The bottom panel \big( Figs. (g, h, i)\big) shows the probability distributions for Run (3).}
\label{fig:a}
\end{minipage}
\end{figure}
\end{center}
\twocolumngrid
\section{Discrete-time Un-biased Quantum Walk (DTUBQW)}
In this section, we generalize the quantum walk by selecting the size of the step not to be the unit size.
\begin{align}\notag
\hat{S}=&\ket 0_C \bra 0 \otimes \sum_x \ket {x-k}_P \bra x
\label{shift} \\+&\ket 1_C \bra 1 \otimes \sum_x \ket {x+l}_P \bra x,
\end{align}
where the parameters $k$ and $l$ are integers.  $k$   determines the size of the step to the right and $l$ to the left.  The walk operator, $\hat{U}=\hat{S}\hat{C}$, acts on the initial state ($\phi_{\text{int}}$) of the walker and leads to the final state of the walk. For DTUBQW k $=$ l and the different step sizes produce different probability distributions.
The stepwise evolution of the initial state $\phi_{\text{int}}$ of the walker can be mathematically written as,

\begin{align}
\notag \ket 0_C \otimes \ket 0_P &\xRightarrow[\text{}]{\text{1st}}\cos\theta\ket 0_C\otimes \ket{{-k}_P}
\\ \notag &+\sin\theta\ket 1_C\otimes \ket{{l}_P}
\\ \notag &\xRightarrow[\text{}]{\text{2nd}}\cos^{2}\theta \:\ket 0_C \otimes \ket{{-2k}_P}
\\ \notag &+\sin^{2}\theta \:\ket 0_C\otimes \ket {{l-k}_P}
\\ \notag &+\cos\theta \sin\theta \ket 1_C \otimes\ket {{l-k}_P}
\\ \notag &-\sin\theta \cos\theta \ket 1_C \otimes\ket {{2l}_P}
\\ \notag  \xRightarrow[\text{}]{\text{Tth}} & \cos^{T}\theta    \: \ket 0_C \otimes\ket {{[-k T]}_P}
\\\label{f1} &+\ldots+ \cos^{T-1}\theta \sin\theta   \: \ket 1_C \otimes\ket {{[l T]}_P}
\end{align}
It is observed that position of a peak depends upon the step size given by the following relation;
\begin{align}
\notag\text{Position  of  the  Peak} = &2 \times \text{Number of the Peak}
\\ &  \times \text{Step  Size}
\end{align}
In Fig. \ref{fig:b} we have shown the probability distribution for DTUBQW for different rotation angles i.e. $\theta=\pi/$, $\theta=\pi/4$ and $\theta=\pi/3$. The probability depends upon the rotation angle as well as on step size. For instance,  for angle $(\theta=\pi/4)$ there are more higher peaks as compared to $(\theta=\pi/6)$ and $(\theta=\pi/3)$.
\onecolumngrid
\begin{center}
\noindent
\begin{figure}[H]
\begin{minipage}{0.98\columnwidth}
\begin{tabular}{ccc}
      \addheight {\includegraphics[width=57mm]{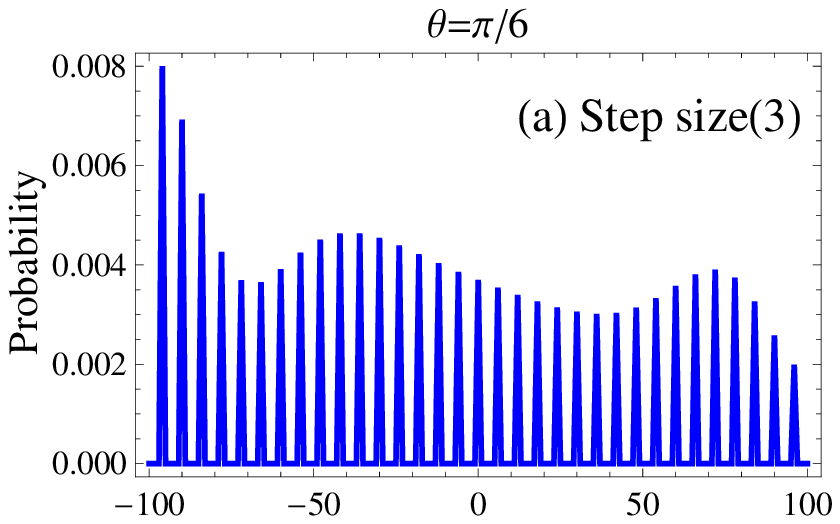}} &
      \addheight {\includegraphics[width=57mm]{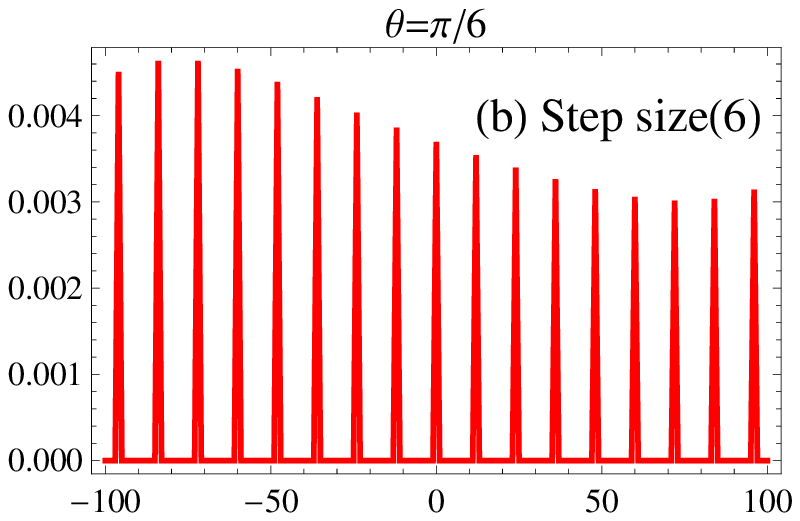}} &
      \addheight{\includegraphics[width=57mm]{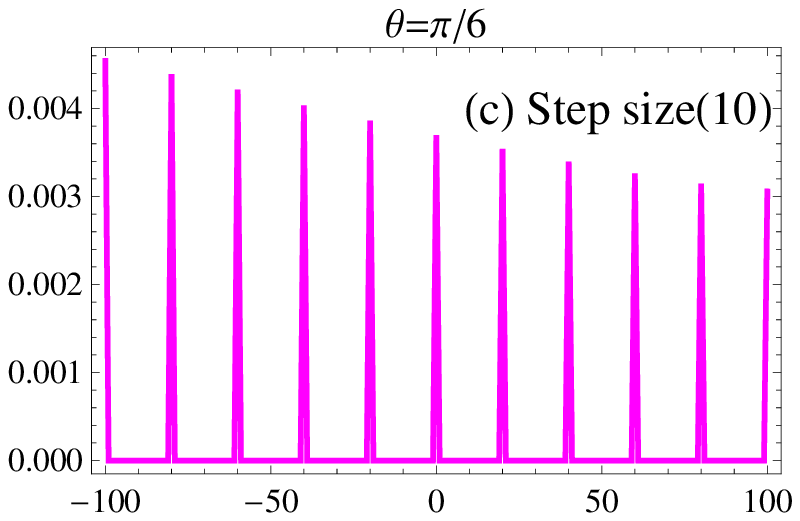}}\\
      \addheight {\includegraphics[width=57mm]{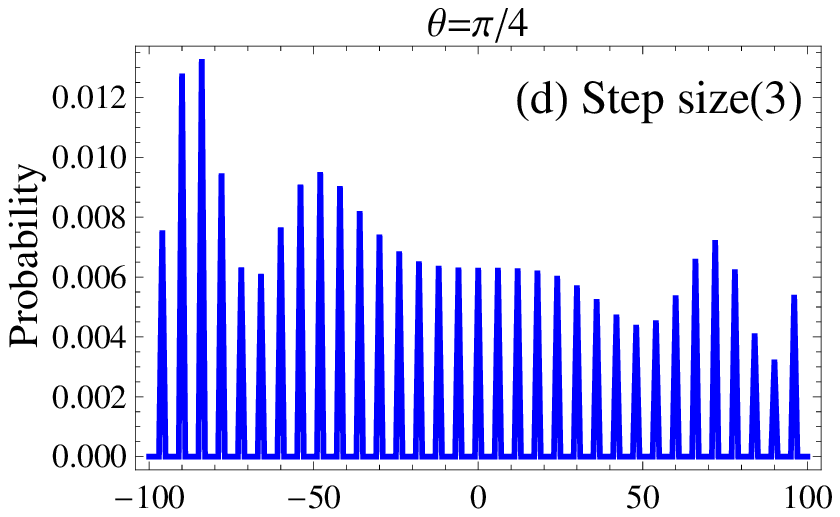}} &
      \addheight {\includegraphics[width=57mm]{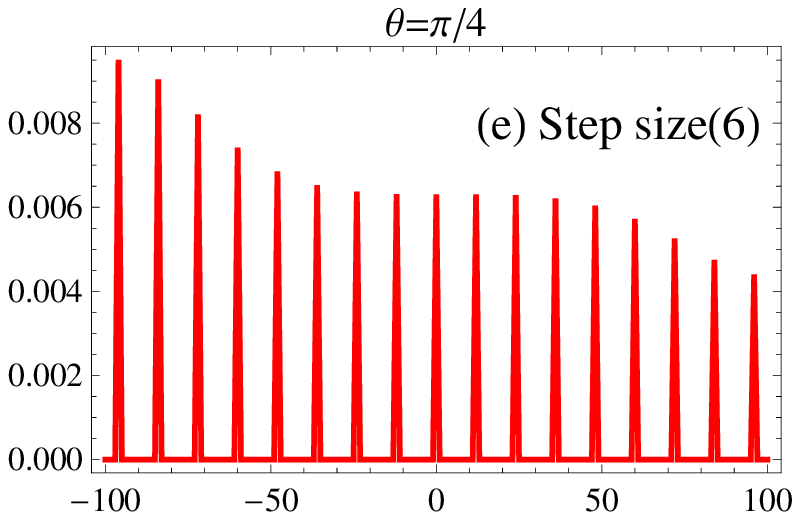}} &
      \addheight{\includegraphics[width=57mm]{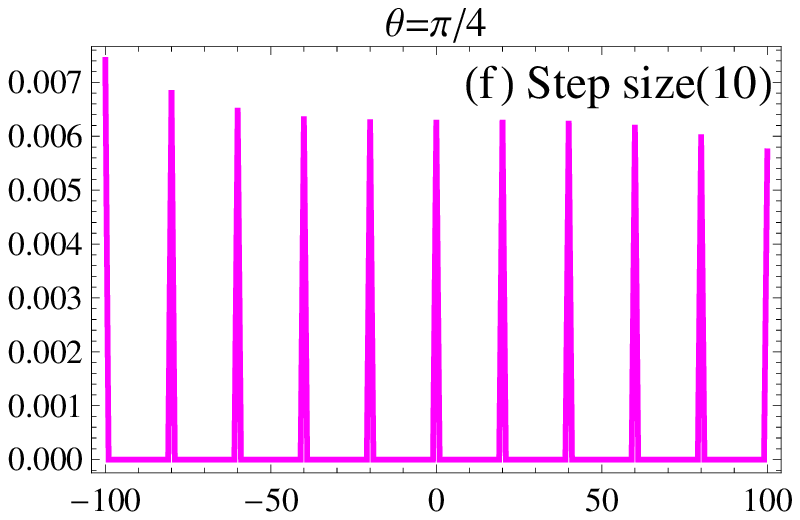}}\\
      \addheight {\includegraphics[width=57mm]{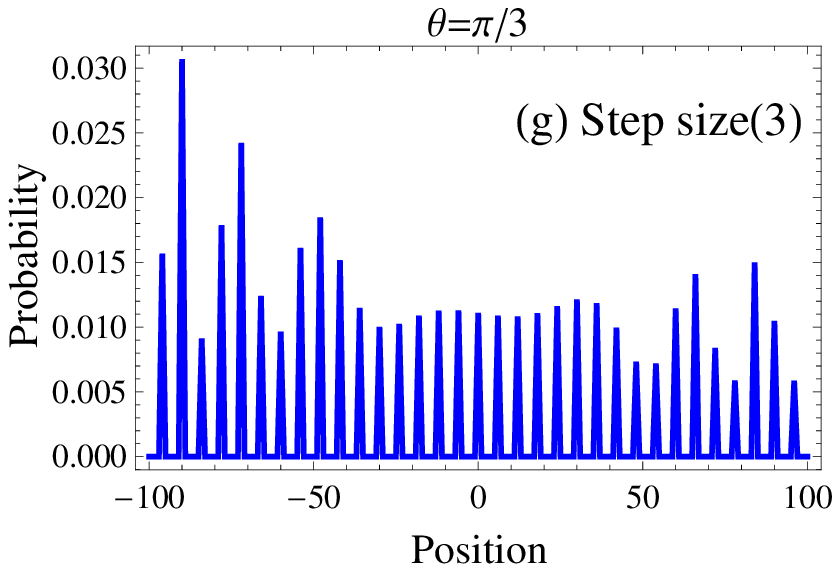}} &
      \addheight {\includegraphics[width=57mm]{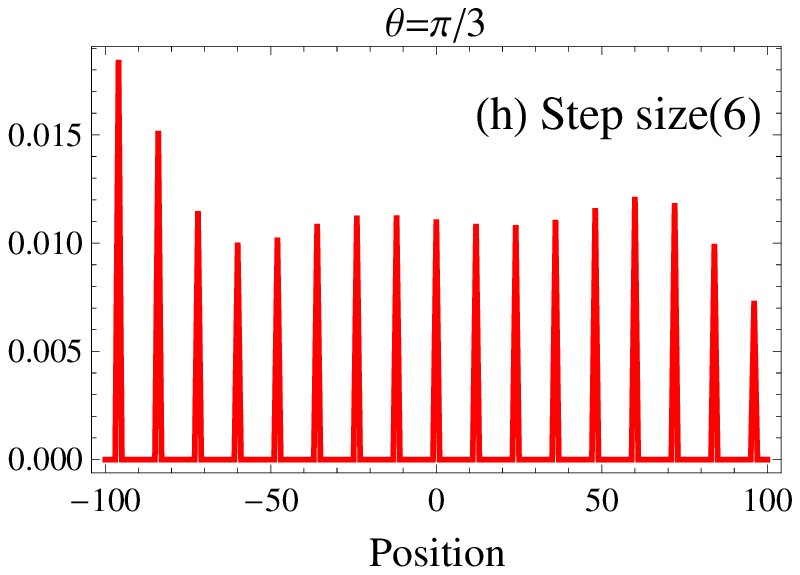}} &
      \addheight{\includegraphics[width=57mm]{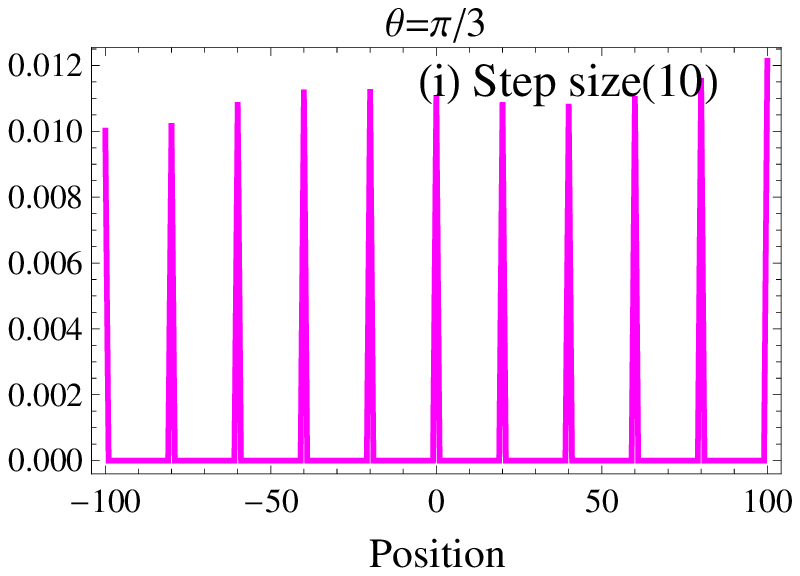}}\\
 \end{tabular}
\caption{Probability distributions of the DTUBQW after T $=100$. In the top panel \big(Figs. (a, b, c)\big)  probability distributions are for step size 3 and rotation angle $(\theta=\pi/6)$. In the middle panel  \big(Figs.(d, e, f)\big)  probability distributions are  for step size 6 and  rotation angle $(\theta=\pi/4)$. In the bottom panel \big(Figs. (g, h, i)\big)  probability distributions are for step size 10 and rotation angle $(\theta=\pi/3)$.}
\label{fig:b}
\end{minipage}
\end{figure}
\end{center}
\twocolumngrid
\section{Discrete-time Biased Quantum Walk (DTBQW)}
In this section, we analyze the biased quantum walk by selecting the un-equal step size to right and left. Fig. \ref{fig:c} depicts the probability distribution of a DTBQW for different rotation angles i.e. $(\theta=\pi/6)$, $(\theta=\pi/4)$ and$(\theta=\pi/3)$.  Comparing Fig. \ref{fig:c} and Fig. \ref{fig:d} it is clear that the probability distribution is asymmetric when we exchange left and right step sizes. The probability depends upon the angle, size of the step as well on the difference between the Left Step Size (LSS) and Right Step Size (RSS).
\onecolumngrid
\begin{center}
\noindent
\begin{figure}[H]
\begin{minipage}{0.98\columnwidth}
\begin{tabular}{ccc}
      \addheight {\includegraphics[width=57mm]{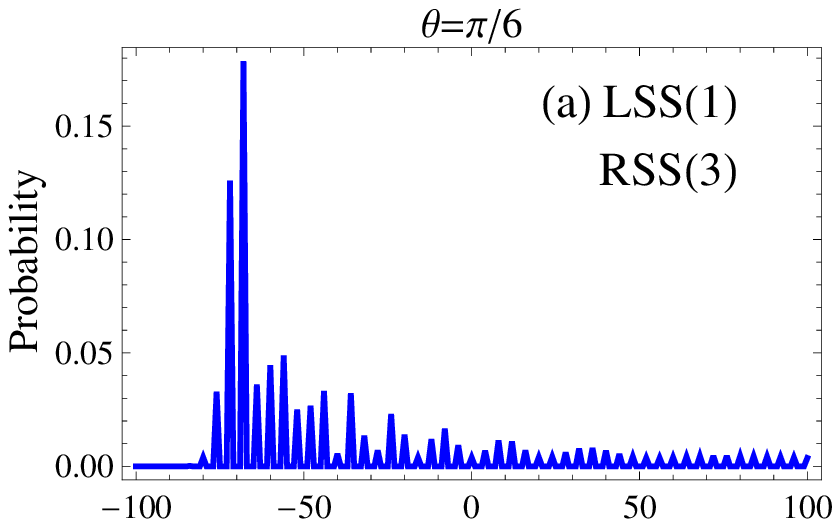}} &
      \addheight {\includegraphics[width=57mm]{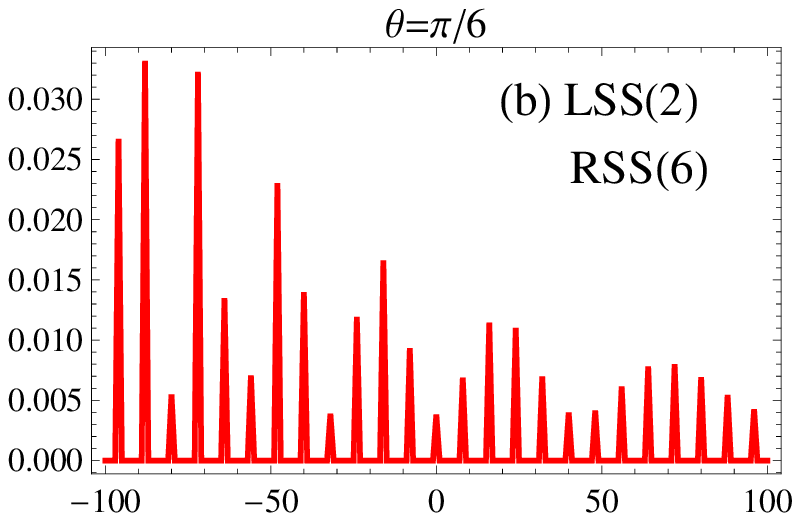}} &
      \addheight{\includegraphics[width=57mm]{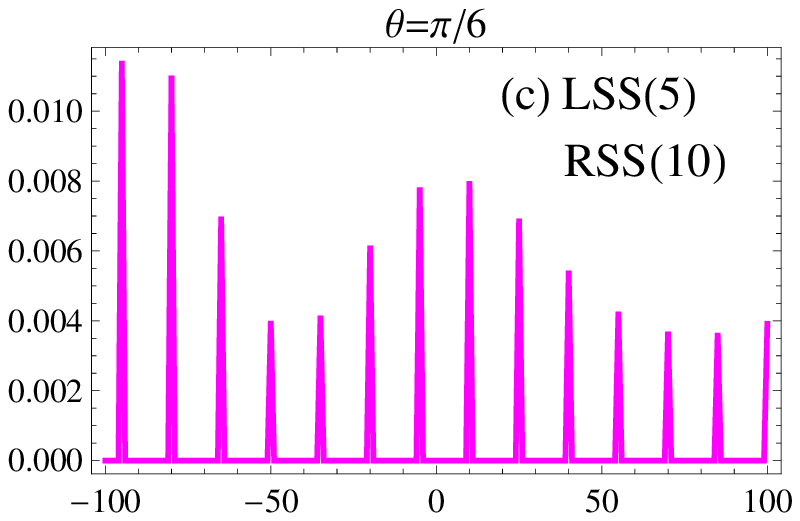}}\\
      \addheight {\includegraphics[width=57mm]{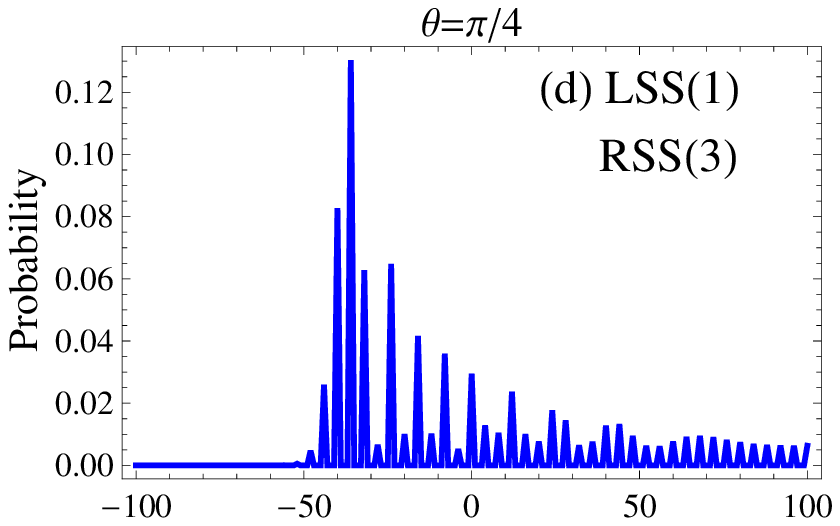}} &
      \addheight {\includegraphics[width=57mm]{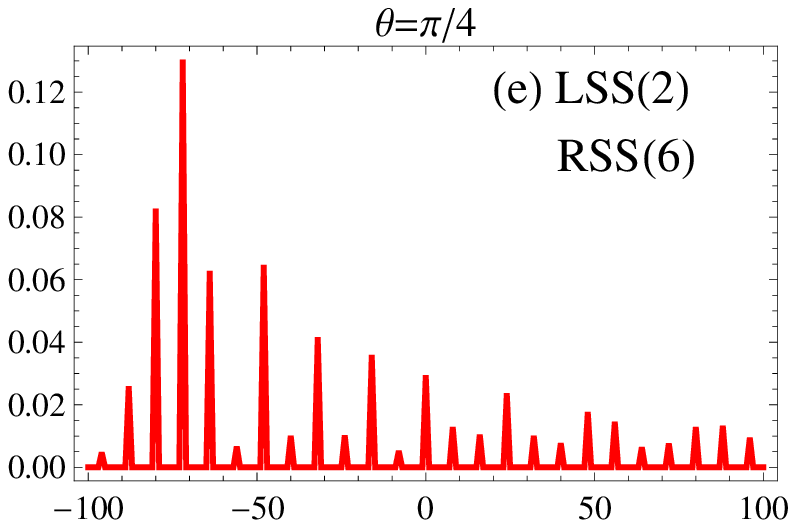}} &
      \addheight{\includegraphics[width=57mm]{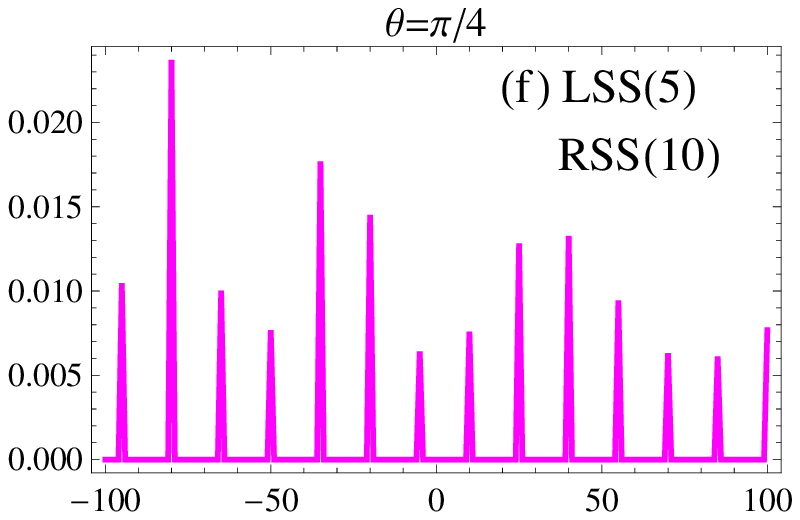}}\\
      \addheight {\includegraphics[width=57mm]{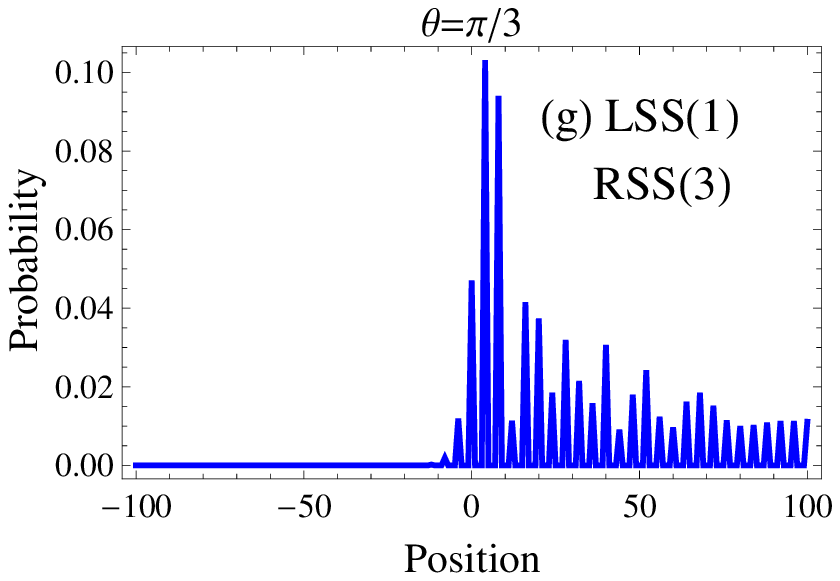}} &
      \addheight {\includegraphics[width=57mm]{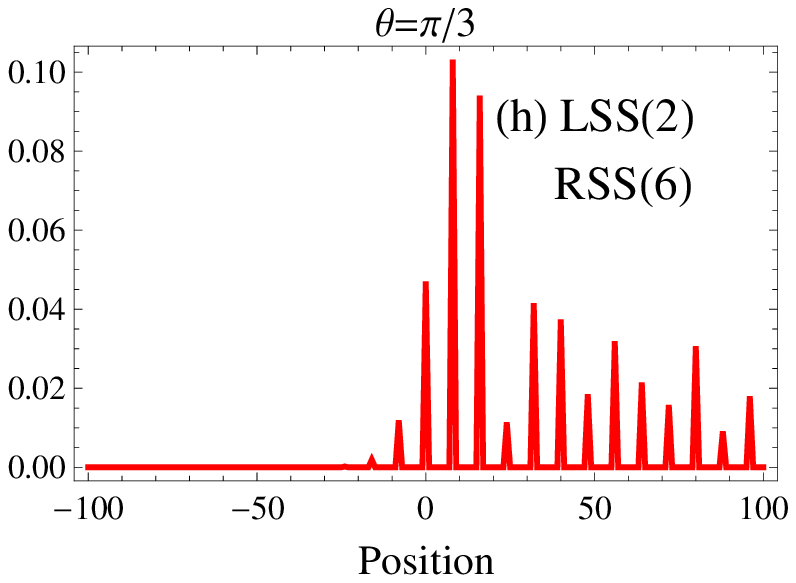}} &
      \addheight{\includegraphics[width=57mm]{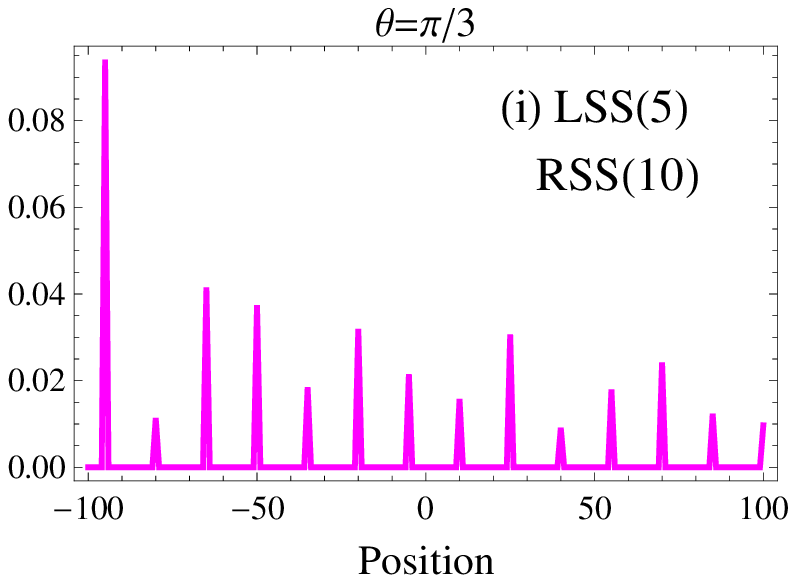}}\\
 \end{tabular}
\caption{Probability distribution of the DTBQW after T $=100$. The Left Step Size (LSS)  is larger than Right Step Size (RSS). In the top panel \big(Figs. (a, b, c)\big)  probability distributions are for rotation angle $(\theta=\pi/6)$. In the middle panel  \big(Figs.(d, e, f)\big)  probability distributions are  for rotation angle $(\theta=\pi/4)$. In the bottom panel \big(Figs. (g, h, i)\big)  probability distributions  for rotation angle $(\theta=\pi/3)$.}
\label{fig:c}
\end{minipage}
\end{figure}
\end{center}
\twocolumngrid
In the DTBQW, probability distributions obtained are quite versatile and one can generate desired probability by selecting the LSS and RSS to one's choice. Such a large number of probabilities available can help to simulate diverse quantum phenomenon.
\onecolumngrid
\begin{center}
\noindent
\begin{figure}[H]
\begin{minipage}{0.98\columnwidth}
\begin{tabular}{ccc}
      \addheight {\includegraphics[width=57mm]{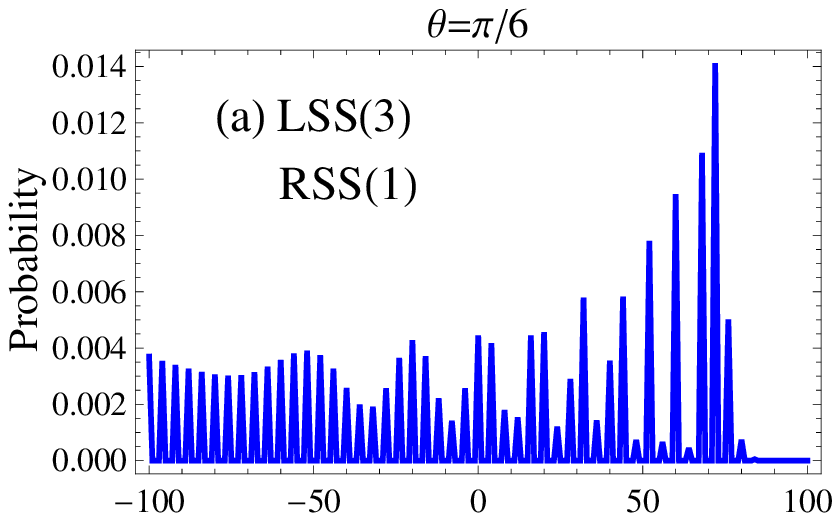}} &
      \addheight {\includegraphics[width=57mm]{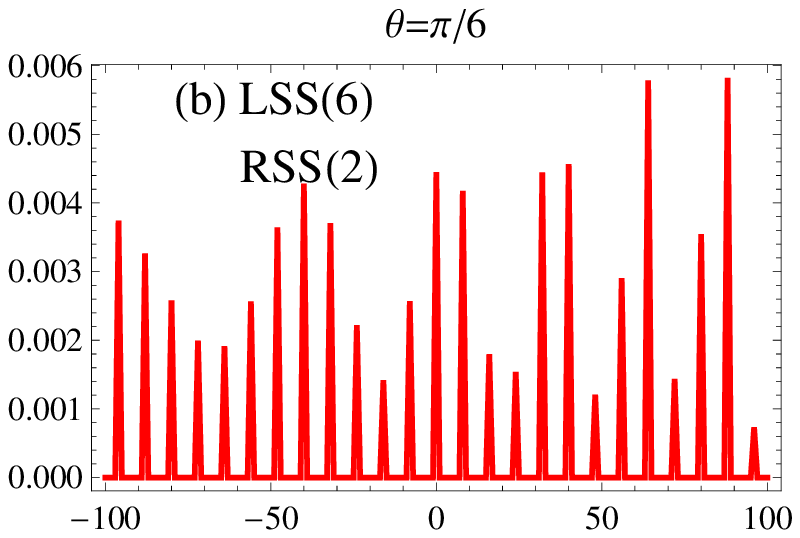}} &
      \addheight{\includegraphics[width=57mm]{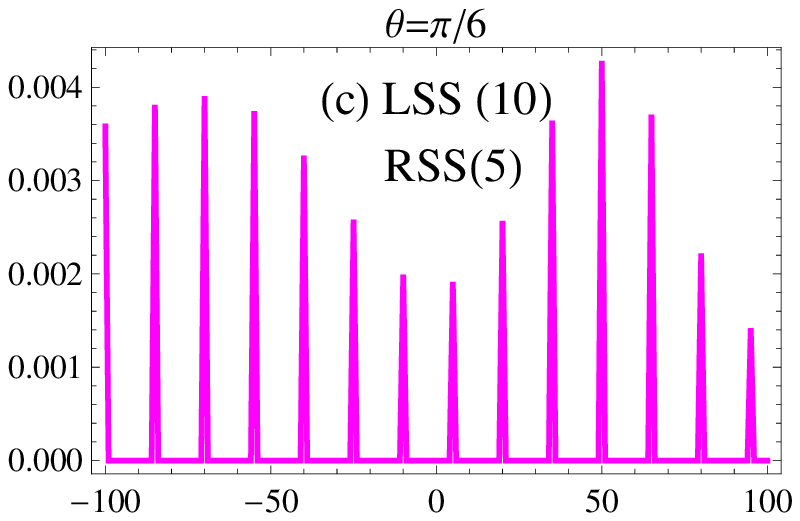}}\\
      \addheight {\includegraphics[width=57mm]{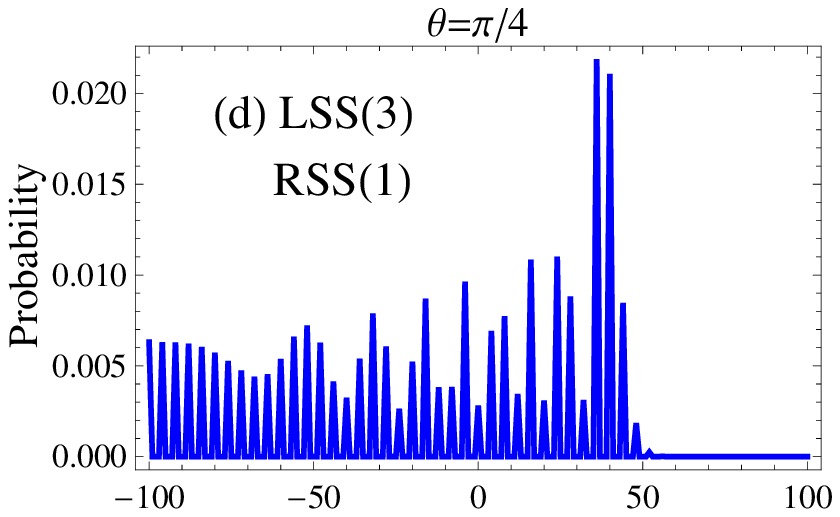}} &
      \addheight {\includegraphics[width=57mm]{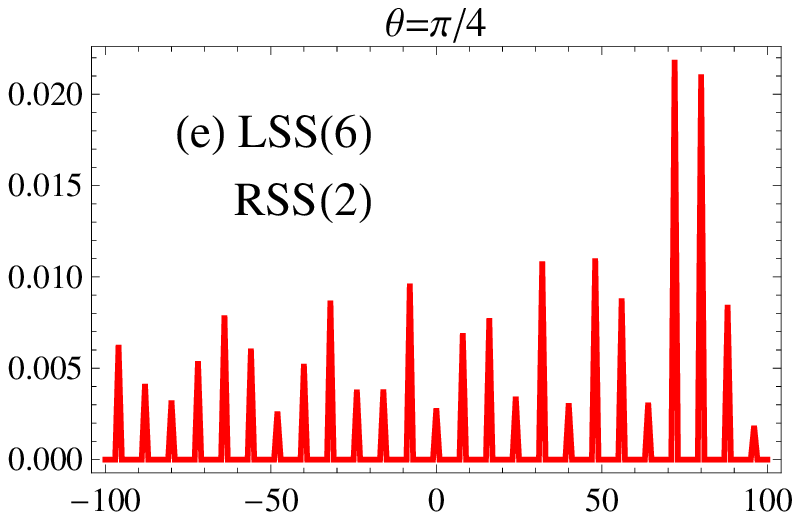}} &
      \addheight{\includegraphics[width=57mm]{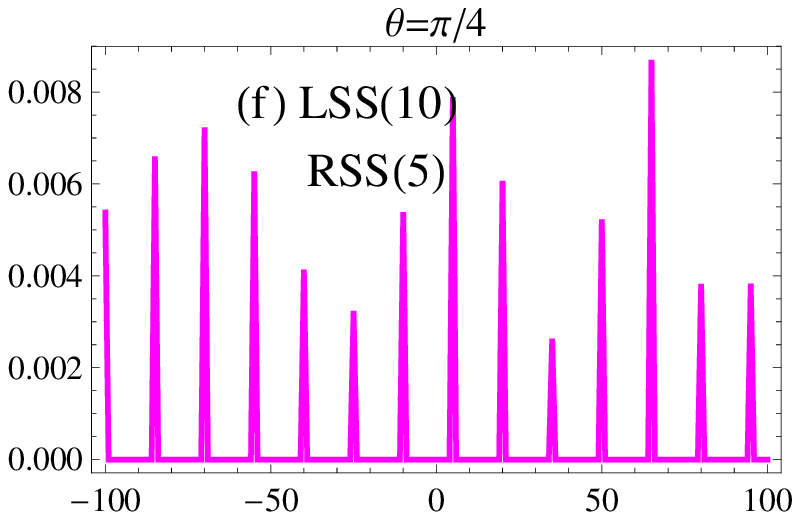}}\\
      \addheight {\includegraphics[width=57mm]{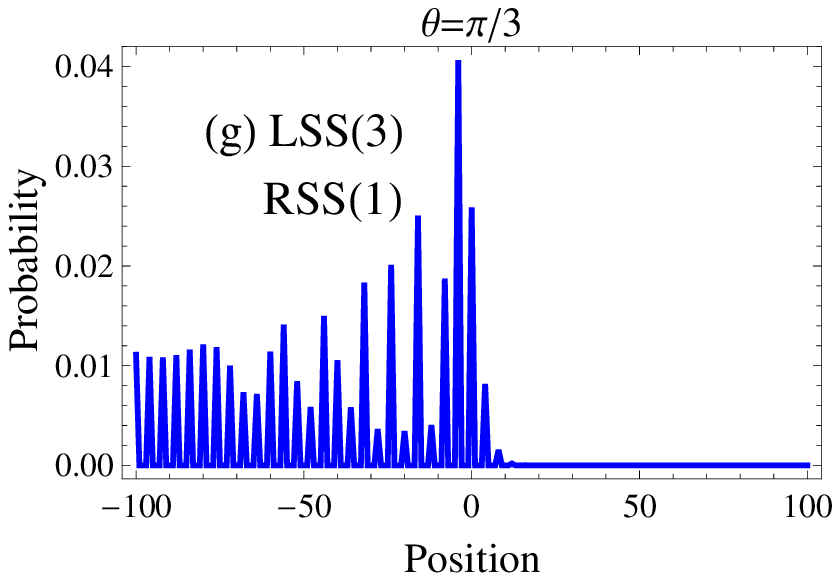}} &
      \addheight {\includegraphics[width=57mm]{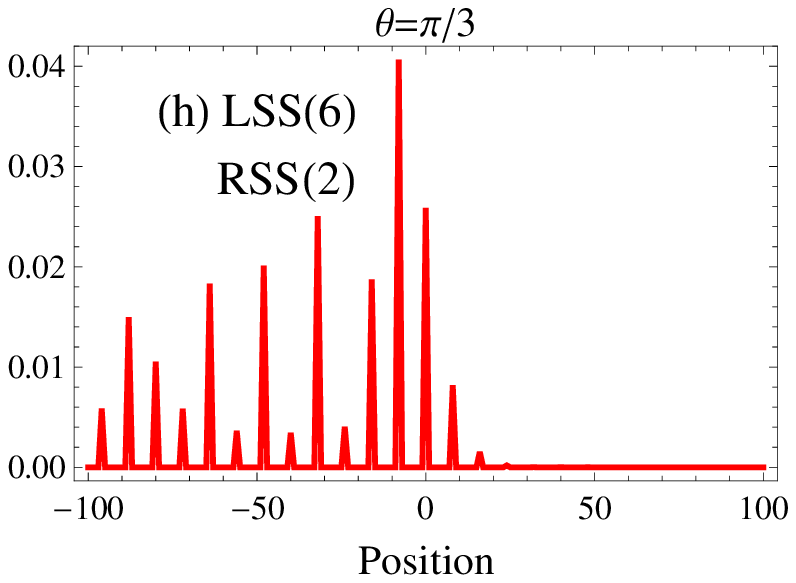}} &
      \addheight{\includegraphics[width=57mm]{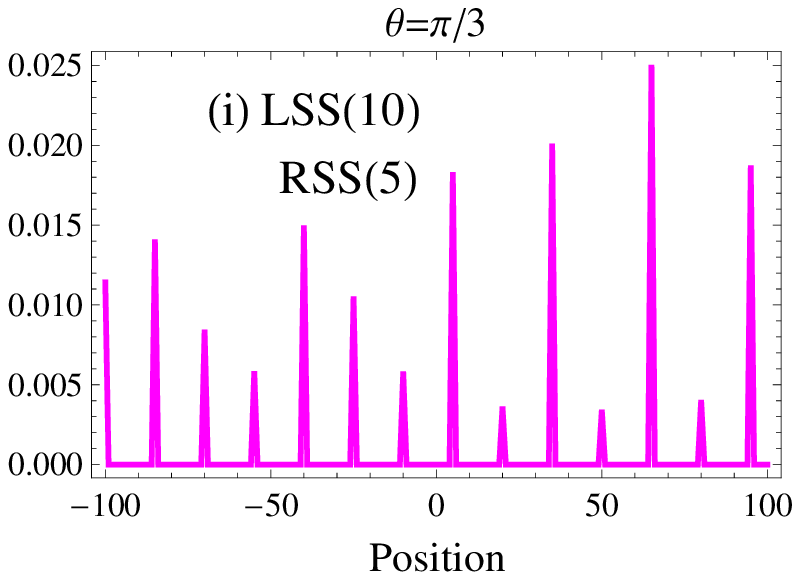}}\\
 \end{tabular}
\caption{Probability distribution of the DTBQW after T $=100$. The Left Step Size (LSS)  is smaller than Right Step Size (RSS). In the top panel \big(Figs. (a, b, c)\big)  probability distributions are for rotation angle $(\theta=\pi/6)$. In the middle panel  \big(Figs.(d, e, f)\big)  probability distributions are  for rotation angle $(\theta=\pi/4)$. In the bottom panel \big(Figs. (g, f, i)\big)  probability distributions are for rotation angle $(\theta=\pi/3)$.}
\label{fig:d}
\end{minipage}
\end{figure}
\end{center}

\twocolumngrid
\section{Standard Deviation} \label{sd}
We calculated the standard deviation ($\sigma$) for all three types of walks. It is observed that generally, they all spread faster than standard DTQW. In Fig. \ref{fig:sd}  (a)  it can be seen that the standard deviation of DTRSQW goes higher than DTQW especially in the case of run-III where the spread is quite fast after 22 steps. It is possible that in some runs spread is even faster. However, in Fig. \ref{fig:sd}  (b) and (c) the standard deviation becomes relatively slow. It indicates that for large intervals spreads become slow and can be always smaller than DTQW in some runs.
\par In Fig.  \ref{fig:sd}  (d), (e), and (f) standard deviations of DTUBQW are shown for rotation angles $\theta = \pi/6$, $\theta = \pi/4$, and $\theta = \pi/3$, respectively. The deviation behavior resembles sawtooth with tooth size getting bigger with rotation angle and step size. For smaller angle and small step size, $\sigma$ jumps higher than $\sigma$ of DTQW, the number of jumps as well as $\sigma$ increases with increasing angle.
\par In Fig. \ref{fig:sd}  (g), (h), and (i) $\sigma$ for DTBQW are plotted for Left Step Size (LSS) that is larger than Right Step Size (RSS). As the step size becomes larger the $\sigma$ decreases  however, it increases with the larger angles.  In Fig. \ref{fig:sd}  (j), (k), and (l) the $\sigma$ for DTBQW for  LSS smaller than RSS are plotted. The $\sigma$ in both cases is not the same and it appears that for larger LSS $\sigma$ is higher.
\onecolumngrid
\begin{center}
\noindent
\begin{figure}[H]
\begin{minipage}{0.98\columnwidth}
\begin{tabular}{ccc}
      \addheight {\includegraphics[width=57mm]{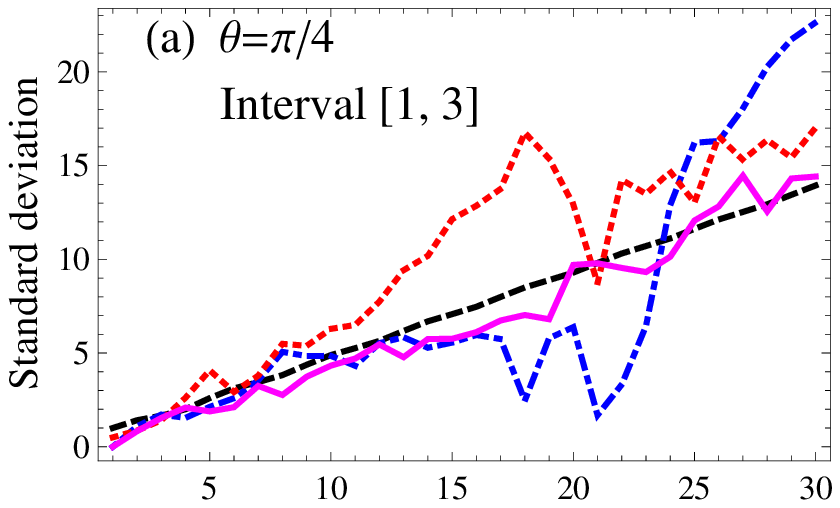}} &
      \addheight {\includegraphics[width=57mm]{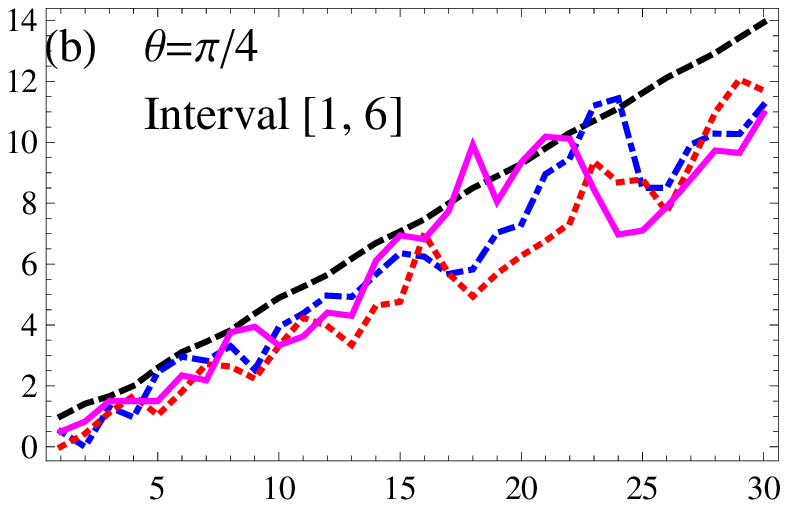}} &
      \addheight{\includegraphics[width=57mm]{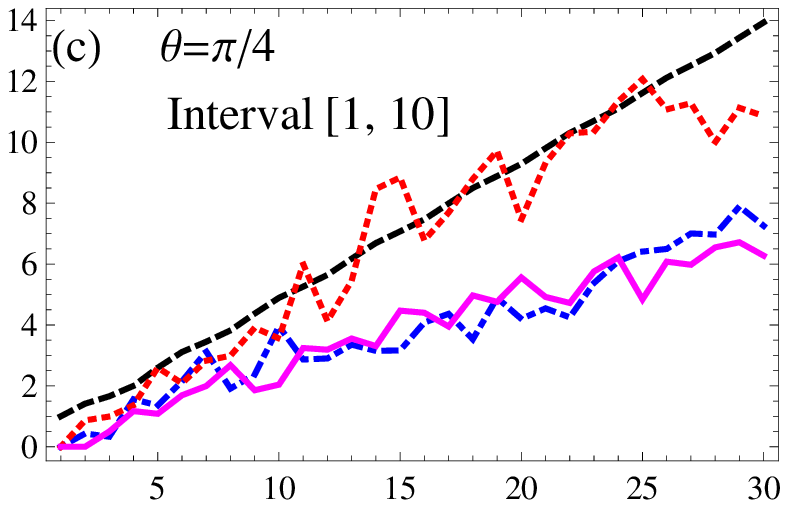}}\\
      \addheight {\includegraphics[width=57mm]{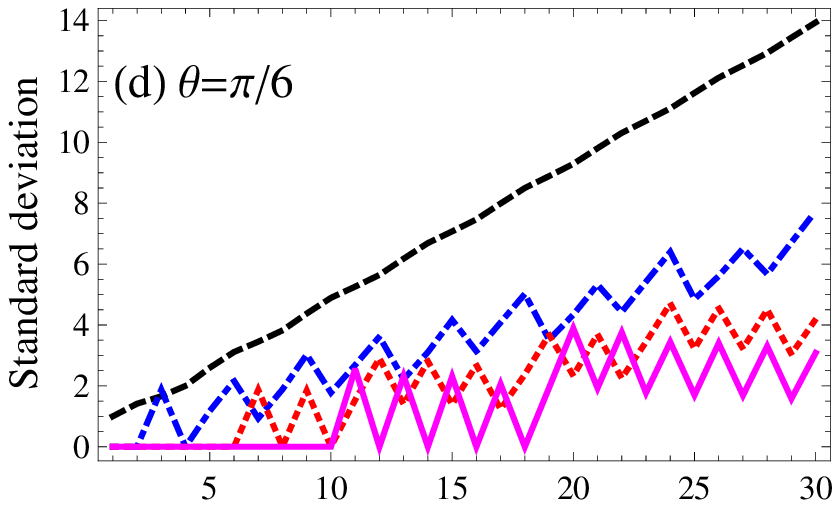}} &
      \addheight {\includegraphics[width=57mm]{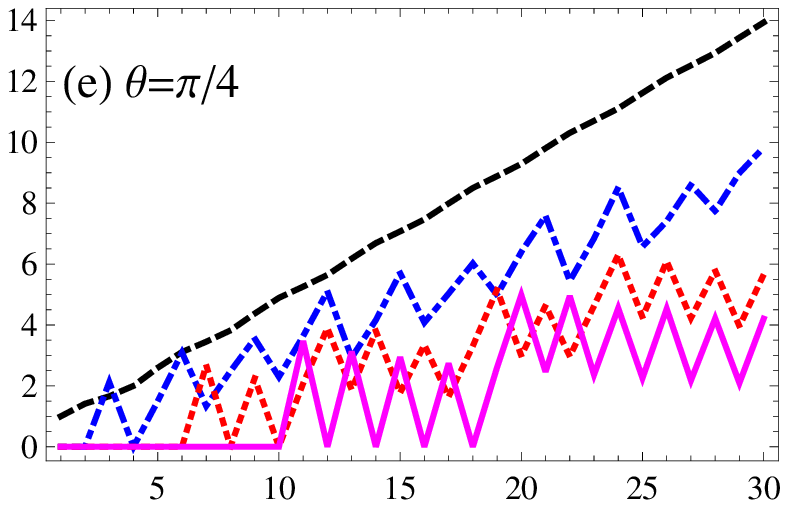}} &
      \addheight{\includegraphics[width=57mm]{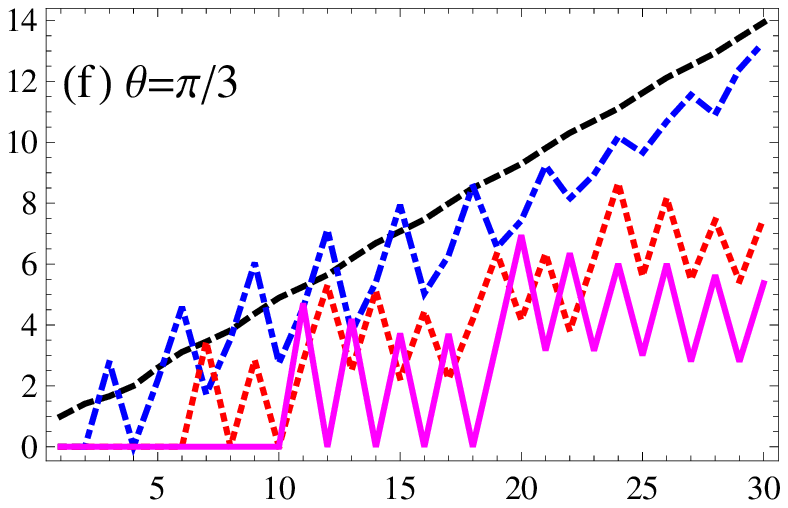}}\\
       \addheight {\includegraphics[width=57mm]{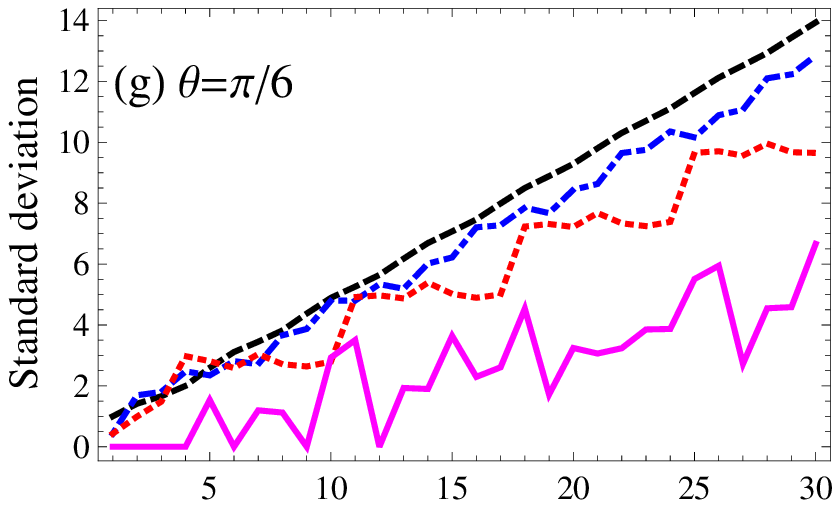}} &
      \addheight {\includegraphics[width=57mm]{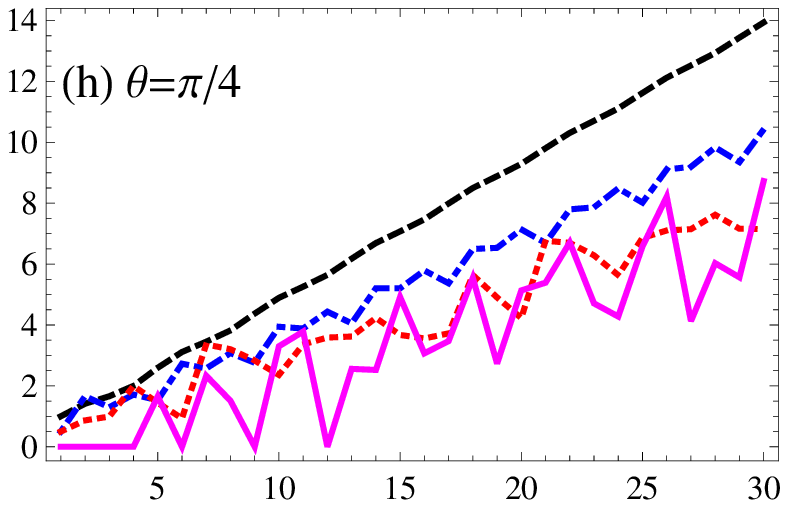}} &
      \addheight{\includegraphics[width=57mm]{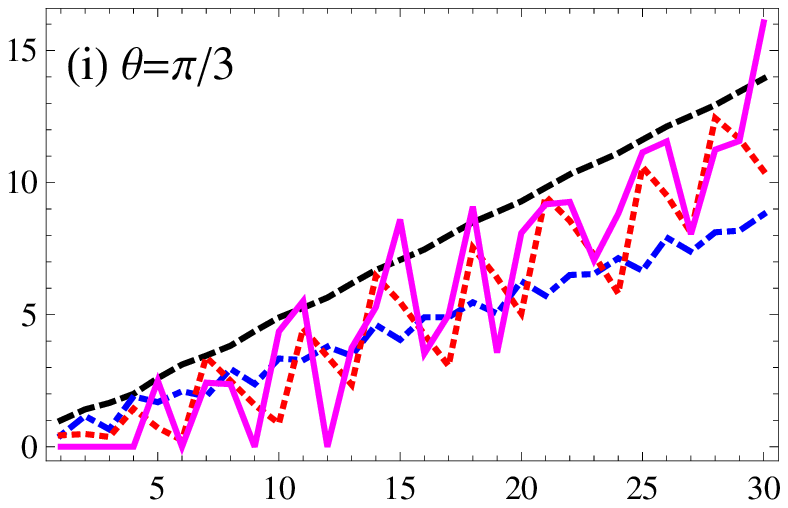}}\\
       \addheight {\includegraphics[width=57mm]{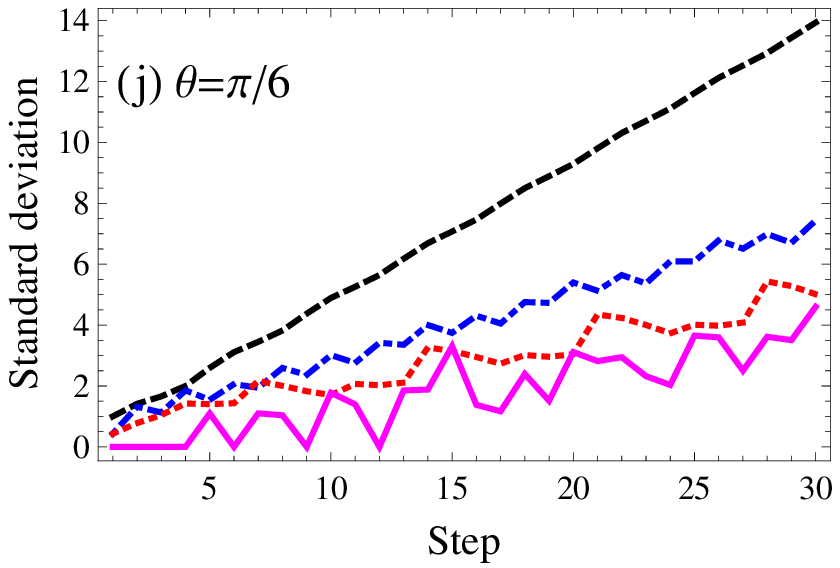}} &
      \addheight {\includegraphics[width=57mm]{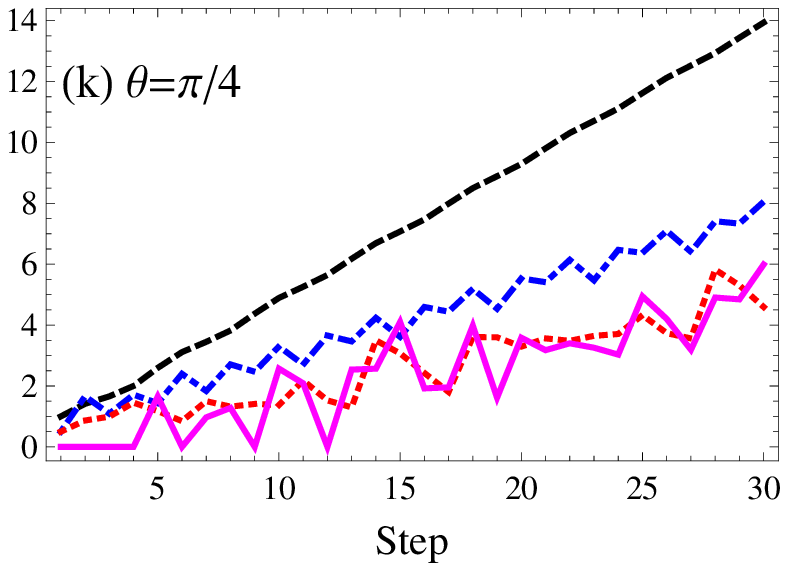}} &
      \addheight{\includegraphics[width=57mm]{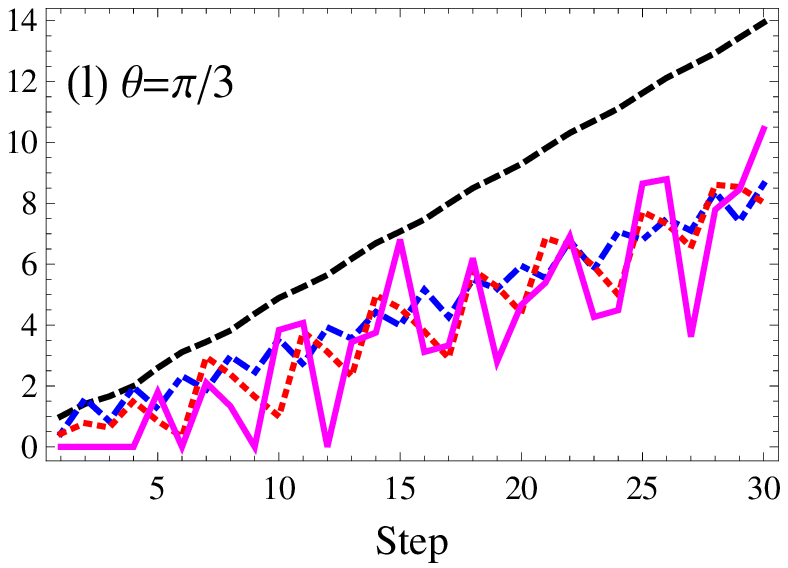}}\\
 \end{tabular}
\caption{Standard deviation ($\sigma$) for T = 30 steps, dashed (black) line shows the $\sigma$ of standard DTQW with unit step size shift. In sub-figures (a), (b) and (c)  dotted-dashed (blue) line represents run-I, dotted (red) line represents run-II, and solid (magenta) line represents run-III of DTRSQW.  While in sub-figures (d), (e) and (f)  dotted-dashed (blue) line represents step size 3, dotted (red) line represents the step size 6, and solid (magenta) line represents step size 10 of DTUBQW. In sub-figures (g), (h) and (i) dotted-dashed (blue) line represents LSS = 3 and RSS = 1, dotted (red) line represents LSS = 6 and RSS = 1, and solid (magenta) line represents LSS = 10 and RSS = 5 of DTBQW. While in sub-figures (j), (k) and (l) dotted-dashed (blue) line represents LSS = 1 and RSS = 3, dotted (red) line represents LSS = 1 and RSS = 6, and solid (magenta) line represents LSS = 5 and RSS = 10 of DTBQW.}
\label{fig:sd}
\end{minipage}
\end{figure}
\end{center}

\twocolumngrid

\section{Conclusion}\label{conc}

Three different types of quantum walks are introduced in this work. The first walk is called DTRSQW, it introduces randomness in quantum walks. The step size is randomly chosen after each step from a specific interval that makes the probability distribution completely random. Its standard deviation is higher than the standard  DTQW and hence it spreads faster than DTQW. This property makes DTRSQW useful for designing new quantum algorithms and simulating random quantum physical phenomenon.  The second walk is called DTUBQW, it gives well-behaved probability distribution. The study was conducted for many angles and step sizes that show that the probability depends upon the angle and step size. The relation between step size and position of peaks has been found to obey the formula. The number of peaks in probability distribution also depends on the rotation angle, we get less number of peaks as the angles go higher. The standard deviation shows the sawtooth behavior that makes this walk another candidate for inventing new quantum algorithms. The third type of walk is called DTBQW, it has different step sizes for the left and right shifts. It gives different probability distributions for the exchange of step sizes from left to right. The standard deviation is higher when Left Step Size (LSS) is larger than the Right Step Size (RSS).  All these walks can be used to simulate different physical phenomena and design new quantum algorithms.

\par
This study of a one-dimensional quantum walks can be generalized to quantum walks with entangled coins and for higher dimensions with single and multi-particles. It can also be generalized to more than two entangled qubits. It will be interesting to use different coins for these walks. We will leave these matters for future work.


\begin{thebibliography}{1}
\bibitem{brownian}F. B. Knight, On the Random Walk and Brownian Motion, \href{https://www.jstor.org/stable/1993657?seq=1#page_scan_tab_contents}{Trans. Amer. Math. Soc. \textbf{103},(1962)}.
\bibitem{diff}S. Hoshino , K. Ichida, Solution of partial differential equations by a modified random walk, \href{https://link.springer.com/article/10.1007/BF01398459}{Numerische Mathematik \textbf{18},1 (1971)}.
\bibitem{slutsky}S. Michael, Protein-DNA interaction, random walks and polymer statistics,\href{ http://hdl.handle.net/1721.1/32295}{Thesis (Ph. D.), Massachusetts Institute of Technology, Dept. of Physics (2005)}.
\bibitem{kempe}J. Kempe, Quantum random walks: An introductory overview, \href{https://www.tandfonline.com/doi/abs/10.1080/00107151031000110776}{Contemp. Phys. \textbf{44}, 307 (2003)}.

\bibitem{andracas}S. E. Venegas-Andraca, Quantum walks: A comprehensive review, \href{https://dl.acm.org/citation.cfm?id=2386759}{Quantum Inf. Proc. \textbf{11}, 1015 (2012)}.
\bibitem{cont1} E. Farhi and S. Gutmann, Quantum computation and decision trees, \href{https://journals.aps.org/pra/abstract/10.1103/PhysRevA.58.915}{Phys. Rev. A \textbf{58}, 915(1998)}.

\bibitem{cont2} Matteo A. C. Rossi, C. Benedetti, M. Borrelli, S. Maniscalco, and Matteo G. A. Paris, Continuous-time quantum walks on spatially correlated noisy lattices, \href{https://journals.aps.org/pra/abstract/10.1103/PhysRevA.96.040301}{Phys. Rev. A \textbf{96}, 040301(2017)}.

\bibitem{cont3}O. Mülken and A. Blumen, Continuous-time quantum walks: Models for coherent transport on complex networks, \href{https://www.sciencedirect.com/science/article/abs/pii/S0370157311000184}{Phys. Rep \textbf{502}, 2-3(2011)}
\bibitem{disc1} N. B. Lovett, S. Cooper, M. Everitt, M. Trevers and V. Kendon, Universal quantum computation using the discrete-time quantum walk, \href{https://journals.aps.org/pra/abstract/10.1103/PhysRevA.81.042330}{Phys. Rev. A \textbf{81}, 042330(2010)}.

\bibitem{disc2}C. S. Hamilton, S. Barkhofen, L. Sansoni, I. Jex and C. Silberhorn, Driven discrete time quantum walks, \href{http://iopscience.iop.org/article/10.1088/1367-2630/18/7/073008/meta}{New J. Phys \textbf{8}, 8 073008(2016)}.

\bibitem{disc3} P. Kurzyński and A. Wójcik, Discrete-time quantum walk approach to state transfer, \href{https://journals.aps.org/pra/abstract/10.1103/PhysRevA.83.062315}{Phys. Rev. A \textbf{18}, 062315(2016)}.

\bibitem{Ambainis} A. Ambainis, E. Bach, A. Nayak, A. Vishwanath, J. Watrous, One-Dimensional Quantum Walks, \href{http://www.math.uwaterloo.ca/~anayak/papers/AmbainisBNVW01.pdf}{J., Proceedings of the 33th STOC (New York, NY: ACM), (2001)}.

\bibitem{search1}A. M. Childs, R. Cleve, E. Deotto, E. Farhi, S. Gutmann and D. A. Spielman, \textit{Exponential algorithmic speedup by a quantum walk}, in Proceedings of the thirty-fifth annual ACM symposium on theory of computing (San Diego, CA, USA —June 09 - 11, 2003 ) p.59-68.
\bibitem{quantcomp}D. P. DiVincenzo, Quantum Computation, \href{http://science.sciencemag.org/content/270/5234/255}{Science \textbf{270}, 5234(1995)}

\bibitem{search2}A. M. Childs and J. Goldstone, Spatial search by quantum walk, \href{https://journals.aps.org/pra/abstract/10.1103/PhysRevA.70.022314}{Phys. Rev. A , \textbf{70},  022314(2004)}.

\bibitem{speedup}D. Poulin, R. Blume-Kohout, R. Laflamme, and H. Ollivier ,Exponential Speedup with a Single Bit of Quantum Information: Measuring the Average Fidelity Decay, \href{https://journals.aps.org/prl/abstract/10.1103/PhysRevLett.92.177906}{Phys. Rev. Lett. \textbf{92}, 177906(2004)}.

\bibitem{algorithm1}A. Ambainis, Quantum walk algorithm for element distinctness, \href{https://epubs.siam.org/doi/abs/10.1137/S0097539705447311}{SIAM J. Comput \textbf{37}, 1(2007)}.

\bibitem{algorithm2}F. Magniez, M. Santha, and M. Szegedy, Quantum algorithms for the triangle problem, \href{https://epubs.siam.org/doi/abs/10.1137/050643684?mobileUi=0}{SIAM J. Comput \textbf{37}, 2(2007)}.

\bibitem{algorithm3}A. Ambainis, Quantum walks and their algorithmic applications, \href{https://www.worldscientific.com/doi/abs/10.1142/S0219749903000383}{International Journal of Quantum Information \textbf{1}, 507(2003)}.

\bibitem{algorithm4}N. Shenvi, J. Kempe and K. B. Whaley, Quantum random-walk search algorithm, \href{https://journals.aps.org/pra/abstract/10.1103/PhysRevA.67.052307}{Phys. Rev. A \textbf{67}, 052307(2003)}.
\bibitem{topo1}T. Kitagawa, M. S. Rudner, E. Berg and E. Demler, Exploring topological phases with quantum walks, \href{https://journals.aps.org/pra/abstract/10.1103/PhysRevA.82.033429}{Phys. Rev. A \textbf{82}, 033429(2010)}.

\bibitem{topo2} H. Obuse, J. K. Asbóth, Y. Nishimura and N. Kawakami, Unveiling hidden topological phases of a one-dimensional Hadamard quantum walk, \href{https://journals.aps.org/prb/abstract/10.1103/PhysRevB.92.045424}{Phys. Rev. B \textbf{92}, 045424(2015)}

\bibitem{topo3}H. Obuse and N. Kawakami ,Topological phases and delocalization of quantum walks in random environments, \href{https://journals.aps.org/prb/abstract/10.1103/PhysRevB.84.195139}{Phys. Rev. B \textbf{84}, 195139(2011)}

\bibitem{chiral}J. K. Asbóth and H. Obuse ,Bulk-boundary correspondence for chiral symmetric quantum walks, \href{https://journals.aps.org/prb/abstract/10.1103/PhysRevB.88.121406}{Phys. Rev. B \textbf{88}, 121406(2013)}.

\bibitem{dim}T.D. Mackay, S. D. Bartlett, L. T. Stephenson and B. C. Sanders, Quantum walks in higher dimensions, \href{http://iopscience.iop.org/article/10.1088/0305-4470/35/12/304/meta}{J. Phys. A: Math. Gen \textbf{32}, 352745(2002)}.

\bibitem{graphs}V. Kendon, Quantum walks on general graphs, \href{https://www.worldscientific.com/doi/abs/10.1142/S0219749906002195}{Int. J. Quantum Inf \textbf{04},5(2006)}.

\bibitem{mcoin}T. A. Brun, H. A. Carteret, and A. Ambainis, Quantum walks driven by many coins, \href{https://journals.aps.org/pra/abstract/10.1103/PhysRevA.67.052317}{Phys. Rev. A \textbf{67}, 052317(2003)}.

\bibitem{mparticle}P. K. Pathak and G. S. Agarwal, Quantum random walk of two photons in separable and entangled states, \href{https://journals.aps.org/pra/abstract/10.1103/PhysRevA.75.032351}{Phys. Rev. A \textbf{75}, 032351(2007)}.

\bibitem{aperiodic}P. Ribeiro, P. Milman and R. Mosseri, Aperiodic Quantum Random Walks, \href{https://journals.aps.org/prl/abstract/10.1103/PhysRevLett.93.190503}{Phys. Rev. A \textbf{93}, 190503(2004)}.

\bibitem{decoherent}T A. Brun, H. A. Carteret, and A. Ambainis, Quantum random walks with decoherent coins, \href{https://journals.aps.org/pra/abstract/10.1103/PhysRevA.67.032304}{Phys. Rev. A \textbf{67}, 032304(2003)}.

\bibitem{history}A. P. Flitney, D. Abbott and N. F. Johnson, Quantum walks with history dependence, \href{http://iopscience.iop.org/article/10.1088/0305-4470/37/30/013/meta}{J. Phys. A: Math. Gen\textbf{37}, 30(2004)}.

\bibitem{biased}M. \v{S}tefa\v{n}\'{a}k, T. Kiss and I. Jex, Recurrence of biased quantum walks on a line, \href{https://iopscience.iop.org/article/10.1088/1367-2630/11/4/043027/meta}{New J. Phys. , \textbf{11},  043027(2009)}.

\bibitem{sequential}Marcelo A. Pires,  and S\'{i}lvio M. Duarte Queir\'{o}s, Quantum walks with sequential aperiodic jumps, \href{https://arxiv.org/pdf/1910.02254.pdf}{arXiv:1910.02254v1 [quant-ph]  (2019)}.

\bibitem{Bera}  M N Bera, A Acin, M Kus, M Mitchell and M Lewenstein, Randomness in Quantum Mechanics: Philosophy, Physics and
    Technology, \href{https://www.ncbi.nlm.nih.gov/pubmed/29105646}{Rep. Prog. Phys. \textbf{80},  124001(2017)}.

\bibitem{Sension}  R. Sension, Quantum path to photosynthesis, \href{https://www.nature.com/articles/446740a?draft=collection#citeas}{Nature  \textbf{446},   740–741 (2007)}.

\bibitem{Hoyer} S. Hoyer, M. Sarovar, and K. B. Whaley, Limits of quantum speedup in photosynthetic light harvesting, \href{https://iopscience.iop.org/article/10.1088/1367-2630/12/6/065041/meta}{New Journal of Physics  \textbf{12},   065041 (2010)}.

\bibitem{entropy}C. E. Shannon, A Mathematical theory of communication, \href{https://onlinelibrary.wiley.com/doi/abs/10.1002/j.1538-7305.1948.tb01338.x}{Bell Syst. Tech. J \textbf{27},3(1948)}

\bibitem{ent1} A. J. Bracken, D. Ellinas and I. Tsohantjis, Pseudo memory effects, majorization and entropy in quantum random walks, \href{http://iopscience.iop.org/article/10.1088/0305-4470/37/8/L02/pdf}{ J. Phys. A: Math. Gen.  \textbf{37}, L91 (2004)}

\bibitem{ent2}C. M. Chandrashekar, R. Srikanth, and R. Laflamme, Optimizing the discrete time quantum walk using a SU(2) coin, \href{https://journals.aps.org/pra/abstract/10.1103/PhysRevA.77.032326}{Phys. Rev. A \textbf{77},  032326(2010)}.



\end{thebibliography}
\end{document}